\documentclass[12pt,preprint]{aastex}

\usepackage{amsmath}
\usepackage{graphicx}
\usepackage{epsfig}
\usepackage{bm}
\usepackage{color}

\newcount\doepsf
\newcommand{\be}{\begin{equation}}
\newcommand{\ee}{\end{equation}}
\newcommand{\bea}{\begin{eqnarray}}
\newcommand{\eea}{\end{eqnarray}}
\newcommand{\bean}{\begin{eqnarray*}}
\newcommand{\eean}{\end{eqnarray*}}

\begin{document}

\title{Numerical experiments on the detailed energy conversion and spectrum studies in a corona current sheet}

\author{Lei Ni\altaffilmark{1,2}, 
               Jun Lin\altaffilmark{1},
               Zhixing Mei \altaffilmark{1},
               and Yan Li \altaffilmark{1}}

\altaffiltext{1}{Yunnan Observatories, Chinese Academy of Sciences, P.O.
Box 110, Kunming, Yunnan 650011, China; leini@ynao.ac.cn}

\altaffiltext{2}{Key Laboratory of Solar Activity, National
Astronomical Observatories, Chinese Academy of Sciences,
Beijing 100012, China; leini@ynao.ac.cn}

\begin{abstract}
In this paper, we study the energy conversion and spectra in a corona current sheet by 2.5-dimensional MHD numerical simulations. Numerical results show that many Petschek-like fine structures with slow-mode shocks mediated by plasmoid instabilities develop during the magnetic reconnection process. The termination shocks can also be formed above the primary magnetic island and at the head of secondary islands. These shocks play important roles in generating thermal energy in a corona current sheet. For a numerical simulation with initial conditions close to the solar corona environment, the ratio of the generated thermal energy to the total dissipated magnetic energy is around $1/5$ before secondary islands appear. After secondary islands appear, the generated thermal energy starts to increase sharply and this ratio can reach a value about $3/5$. In an environment with a relatively lower plasma density and plasma $\beta$, the plasma can be heated to a much higher temperature. After secondary islands appear, the one dimensional energy spectra along the current sheet do not behave as a simple power law and  the spectrum index increases with the wave number. The average spectrum index for the magnetic energy spectrum along the current sheet is about $1.8$. The two dimensional spectra intuitively show that part of the high energy is cascaded to large $kx$ and $ky$ space after secondary islands appear. The plasmoid distribution function calculated from numerical simulations behaves as a power law closer to $f(\psi) \sim \psi^{-1}$ in the intermediate $\psi$ regime. By using $\eta_{eff} = v_{inflow}\cdot L$, the effective magnetic diffusivity is estimated about $10^{11}\sim10^{12}$~m$^2$\,s$^{-1}$.
\end{abstract}


\section{INTRODUCTION}
\label{sec:introduction}
An elongated current sheet attached above the flare loop top \cite[]{2003ApJ...596L.251S, 2005ApJ...622.1251L, 2010ApJ...723L..28L} is usually observed in an eruptive solar flare. Magnetic reconnection inside current sheets(CSs) plays an important role to release and convert the magnetic energy to the plasma thermal and kinetic energy. Reconnection inflows and high speed outflows in CSs have been recognized by lots of observations \cite[]{2007ApJ...661L.207W, 2012ApJ...745L...6T, 2012ApJ...754...13S}. The speed of  outflows ranges from  $100-1000$ ~km\,s$^{-1}$. Many studies show that  the outflow speed is around $100-450$~km\,s$^{-1}$ \cite[]{2004ApJ...605L..77A, 2011ApJ...730...98S,  2012ApJ...745L...6T, 2013ApJ...767..168L, 2014A&A...567A..11Z}. The maximum outflow velocity can reach $1000$~km\,s$^{-1}$ \cite[]{2003SoPh..217..267I, 2013ApJ...767..168L}, which is close to the Alfv\'en velocity in solar corona. The temperature of the plasmas relating with a magnetic reconnection process can be heated to around 3~MK $\sim$ 20~MK. The 3~MK plasma was observed above the post-flare loops by \cite{2012ApJ...751...21L} in EUV with Hinode/Extreme-Ultraviolet Imaging Spectrometer spectra; Using the Solar and Heliospheric Observatory(SOHO) Ultraviolet Coronagraph Spectrometer (UVCS)  data, \cite{2008ApJ...686.1372C} observed plasmas with the temperature around 6~MK inside the current sheet region; The bright blob with hot plasma in the Atmospheric Imaging Assembly (AIA) $131$\AA ~passband has been seen with the peak temperature $\sim$11 MK \cite[]{2011ApJ...732L..25C}; Reuven Ramaty High Energy Solar Spectroscopic Imager(RHESSI) X-ray spectra and images simultaneously show that the plasma has been heated to $>$ 10~MK in a solar flare in the paper by \cite{2013NatPh...9..489S}; \cite{2013ApJ...777...93S} even observed the hot loop plasmas with the peak temperature reaching $\sim$22~MK by RHESSI  hard X-Ray spectra. In the paper by \cite{2014ApJ...786...73S}, the RHESSI hard X-Ray emissions in Figure 6(b) indicate that some local heating is located at the reconnection site. The local heating regions possibly correspond to some fine structures like magnetic islands within the current sheet. They also predict that the sharp change of temperature across the current sheet in Figure 7(f) in their paper is probably a signature of a slow shock. The generation mechanism of the high temperature emission observed by solar telescopes is still not understood well. \cite{2013ApJ...777...93S} proposed that both the petschek-like reconnection and turbulent reconnection are possible to explain the hot CS plasma.

The Petschek-like reconnection model can be used to explain some observation evidences of white-light, UV and X-ray emissions\cite[]{2010ApJ...722..625K}. However, a nonuniform resistivity which should be enhanced at the X-point is required to produce steady-state Petschek reconnection(see Kulsrud 2001 and references). The physical mechanism about the origin of the enhanced resistivity is still unknown. \cite{2009ApJ...695.1151B} studied the energy conversion mechanism by three-dimensional MHD magnetic reconnection model, they concluded that Petschek reconnection accelerates plasma to convert magnetic energy to bulk kinetic energy,  then the accelerated plasma becomes slowed down and the bulk kinetic energy is transformed to heat.  The hottest plasma can reach around 40~MK in their simulation. However, they assumed a non-uniform resistivity, and the corresponding Lundquist number at  the X-point is only 200, which is much smaller than that in solar corona ($\gtrsim 10^{12}$). Both radiative cooling and heat conduction effect are not included in their model.
 
 Fast magnetic reconnection develops in a turbulent plasma is expected to  produce hot plasma as observed by UCVS and X-ray telescope images\cite[]{2008ApJ...689..572B}. Several types of reconnection models in turbulent plasma are proposed,  e.g. stochastic reconnection\cite[]{1999ApJ...517..700L}, fractal reconnection\cite[]{2001EP&S...53..473S} or  plasmoid instabilities\cite[]{2009PhPl...16k2102B}. These models have been studied analytically and numerically in the past few years, and  some common charicteristics were found. In a high Lundquist number environment, plenty of observations \cite[]{2008JGRA..11311107L, 2010ApJ...713.1292M, 2012ApJ...745L...6T, 2013MNRAS.434.1309L, 2013A&A...557A.115K}, numerical simulations\cite[]{2009PhRvL.103j5004S, 2009PhPl...16k2102B, 2012MNRAS.425.2824M, 2012ApJ...758...20N} and even laboratory experiments\cite[]{2012PhRvL.108u5001D} show that multiple levels of plasmoids can always occur in an unstable magnetic reconnection process. Numerical simulations have demonstrated that the reconnection rate $\gamma$ can be strongly increased to a high value($\sim 0.01$)\cite[]{2010PhPl...17f2104H} after multiple levels of plasmoid instabilities appear, and $\gamma$ weakly depends on Lundquist number.  This value is very close to the reconnection rate measured by observations\cite[]{2010ApJ...722..625K}. However,  few numerical simulations focus on the generation of hot plasma inside the current sheet during plasmoid instabilities. In the paper by \cite{2012ApJ...758...20N}, the plasma densityis ($\sim 10^{18}/m^3$ ) is set much higher than that in the corona environment, and the temperature can only be increased from 1~MK to around 2~MK. 
 
The spectrum studies are important for understanding the energy conversion mechanism and dynamics of plasmoids during the plasmoid instabilities. In the papers by \cite{2012PhPl...19g2902N, 2013PhPl...20f1206N, 2013PhPl...20g2114S, 2011ApJ...737...24B}, one dimensional magnetic and kinetic energy spectra along the current sheet have been studied detailedly. The results show that the spectrum is not a single power law form for both magnetic and kinetic energy, the spectral index increases with the wave number after secondary islands appear. The average spectral index for magnetic energy is around 2.0 and it is around 3.0 for kinetic energy in the previous studies \cite[]{2012PhPl...19g2902N, 2013PhPl...20f1206N}. These values are larger than the Kolmogorov spectral index ($\sim 5/3$). The plasmoid distribution function has also been studied detailedly in some previous papers, e.g. \cite[]{2010PhRvL.105w5002U, 2010PhPl...17a0702F, 2012PhPl...19d2303L, 2012PhRvL.109z5002H, 2013PhPl...20g2114S, 2013ApJ...771L..14G}. Their results also demonstrate that the spectrum of the distribution function does not behave as a single power law.  

In this work, we study the energy conversion during plasmoid instabilities in the solar corona current sheet. The physical parameters such as the initial temperature, density and strength of magnetic field in the simulations are close to those in real corona environment. The evolutions of the reconnection rate, temperature and velocity of the plasma inside the current sheet region have been analyzed. Though, the energy conversion process in a flare current sheet has already been studied detailedly in some previous papers (e.g., Reeves et al. 2010). Compared with these previous studies, a more realistic temperature dependent high Lundquist number ($\gtrsim 10^6$) and much higher resolutions are used in our models. Therefore, we can discover many fine structures related to the heating mechanism inside the current sheet region. Both the one and two dimensional spectra have also been studied carefully in this work. In section II, we present the numerical approach and initial states of our simulations. The results are presented in section III. We summarize our results and give discussions in section IV.   

\section{FRAMEWORK OF NUMERICAL MODELS}
\label{sec:models}
 The 2.5-dimensional one fluid MHD model is used in this work. We only consider the fully ionized hydrogen gas and the plasma are composed with electrons and ions. The temperature of the two species are considered the same ($T_i=T_e=T$).  The MHD equations in our simulations are given by:
 \begin{equation}
 \partial_t \rho = -\nabla \cdot (\rho \textbf{v}),
\end{equation}

\begin{eqnarray}
 \partial_t e &=& -\nabla \cdot [(e+p+\frac {1}{2\mu_0 }\vert \textbf{B}
 \vert^2)\textbf{v}-\frac{1}{\mu_0}(\textbf{v} \cdot \textbf{B})\textbf{B}]  \nonumber\\
  & &+ \nabla \cdot [\textbf{v}\vec{\vec{\tau}}+\frac{\eta}{\mu_0}
  \textbf{B} \times (\nabla \times \textbf{B})] + L_{rad}+H,
\end{eqnarray}

\begin{equation}
  \partial_t (\rho \textbf{v}) = -\nabla \cdot [\rho \textbf{v}
  \textbf{v}
 +(p+\frac {1}{2\mu_0} \vert \textbf{B} \vert^2)I -\frac{1}{\mu_0} \textbf{B} \textbf{B} ]+\nabla \cdot \vec{\vec{\tau}},
\end{equation}

\begin{equation}
  \partial_t \textbf{B} = \nabla \times (\textbf{v} \times \textbf{B}-
  \eta\nabla \times \textbf {B}),
\end{equation}
 
 \begin{equation}
   e = \frac{p}{\Gamma_0-1}+\frac{1}{2}\rho \textbf{v}^2+\frac{1}{2\mu_0}\textbf{B}^2,
 \end{equation}
   
 \begin{equation}
     p = \frac{2 \rho k_BT}{m_i},
 \end{equation}
 
 \begin{equation}
    \vec{\vec{\tau}} = \nu(\nabla \textbf{v}+(\nabla \textbf{v})^{\top}-\frac{2}{3}(\nabla \cdot \textbf{v})\vec{\vec{I}})
 \end{equation}
where $\rho$ is the plasma mass density, $\textbf{v}$ is the centre of mass velocity, $e$ is the total energy density, $\textbf{B}$ is the magnetic field, $\vec{\vec{\tau}}$ is the stress tensor, $m_i$ is the mass of a hydrogen ion, $\eta$ is the magnetic diffusivity, $\nu$ is the dynamic viscosity coefficient and $p$ is plasma thermal pressure. The ratio of specific heats $\Gamma_0$ is set to $5/3$ (ideal gas). The magnetic permeability coefficient $\mu_0$ is set to $4\pi \times 10^{-7}$.  The radiative cooling $L_{rad}$ and heating $H$ for fully ionized high temperature plasma are analytically assumed as\cite[]{1980SoPh...68..351N}:  
 
 \begin{displaymath}
  L_{rad} = \left\{ \begin{array}{ll}
   2.23872 \times 10^{-27} (\frac{\rho}{m_i})^2 T^{-1.385} & \textrm{$2.5 \times 10^{5}$~K $< T <$ $10^{6}$~K}\\
   4.64515 \times 10^{-32} (\frac{\rho}{m_i})^2 T^{-0.604} & \textrm{ $10^{6}$~K $\leq T \leq$ $2\times 10^{7}$~K}\\
   1.73380 \times 10^{-39} (\frac{\rho}{m_i})^2 T^{0.413} & \textrm{ $2\times10^{7}$~K}
\end{array} \right.
\end{displaymath}
 
 \begin{equation}
  H = 4.64515 \times 10^{-32} (\frac{\rho}{m_i})^2 T_0^{-0.604}
 \end{equation}
where $T_0=10^6$~K is the initial temperature in the whole simulation domain at $t=0$. 
 
The simulation domain extends from $x=0$ to $x=L_0$ in $x$ direction and from $y=0$ to $y=2L_0$ in $y$ direction, with $L_0=10^8$~m. Open boundary conditions are used in both $x$ and $y$ directions. 

We use a force-free current sheet with a strong guide field in the center as the initial equilibrium distributions of magnetic fields:
\begin{eqnarray}
  B_{x0}&=&0                              \\
  B_{y0}&=&b_0\tanh(\frac{x-0.5L_0}{0.05L_0})    \\
  B_{z0}&=&b_0/\cosh(\frac{x-0.5L_0}{0.05L_0}),
\end{eqnarray}
where $b_0=0.001$~T. The initial current sheet width thus is $\delta_0=0.1L_0$. Due to the force-freeness and neglect of gravity, the initial equilibrium thermal pressure and plasma $\beta$ is uniform.  In this work, we have simulated two cases, $\beta=0.1$ in case A and $\beta=0.05$ in case B. These yield the initial plasma thermal pressure $p_0 = 1/(8\pi)$~Pa in case A and $p_0 = 1/(16\pi)$~Pa in case B. Since the initial equilibrium temperature is $T_0=10^6$~K in both of the two cases, we can find that the initial plasma density $\rho_0 \simeq 2.4\times10^{-12}$~kg\,m$^{-3}$ and Alfv\'en velocity $v_{A0}\simeq 580$~km\,s$^{-1}$ in case A, $\rho_0 \simeq 1.2\times10^{-12}$~kg\,m$^{-3}$ and $v_{A0} \simeq 820 $~km\,s$^{-1}$ in case B. Therefore, case A could represent a lower height in solar corona than that for case B. 

The same magnitude ($\sim 0.1$) of initial perturbations of both magnetic field  and velocity are applied at $t=0$ in the two cases to trigger the reconnection process. The forms of perturbations are listed as below:
 \begin{eqnarray}
   b_{xpert} &=& -pert\cdot b_0 \cdot cos(\frac{2\pi x}{L_0}) \cdot sin[\frac{2\pi (y-0.5L_0)}{L_0}] \\
   b_{ypert} &=&  pert\cdot b_0 \cdot sin(\frac{2\pi x}{L_0}) \cdot cos[\frac{2\pi (y-0.5L_0)}{L_0}]\\   
    v_{xpert} &=&  pert\cdot v_{A0} \cdot sin(\frac{2\pi x}{L_0}) \cdot sin(\frac{\pi y}{2L_0}) \\
    v_{ypert} &=& -pert\cdot v_{A0} \cdot sin(\frac{8\pi y}{L_0})\cdot \frac{random_n} {Max(\lvert random_n\rvert)},    
 \end{eqnarray} 
where $pert=0.1$, $v_{A0}$ is the initial Alfv\'en velocity as presented in the above paragraph, $random_n$ is the random noise function in our code,  and $Max(\lvert random_n\rvert)$ is the maximum of the absolute value of the random noise function. The perturbations result in a thinning of the current sheet in two sections between a set of three primary islands, whose midpoints are located at $y=0$, $L_0$, and $2L_0$ (see Figure 1). In this paper, we will focus only on the section in the domain $L_0<y<2L_0$, i.e., the bottom half of the box is used as an auxiliary part of the computation only. Its function is to generate a stationary primary plasmoid at the bottom of the height range of interest, which is not influenced by any effects of a numerical boundary; the primary plasmoid acts like a line-tied bottom of the current sheet of interest.
  
The magnetic diffusivity in the two cases are both assumed as $\eta = 8\times10^7(10^6/T)^{1.5}$. Since the initial temperature is $10^6$~K,  the initial magnetic diffusion is calculated as $\eta_0=8\times10^7$~m$^2$\,s$^{-1}$. The Lundquist number based on this magnetic diffusivity, $v_{A0}$, and $L_0$, which corresponds to the 'global scale' of the current sheet in the upper part of the box, is $S_0=  7.2 \times 10^5$ in case A and $S_0=1.0 \times 10^6$ in case B. As the plasmoid instabilities develop, the highest temperature at the main $X$-point can reach around $10$~MK in our simulations in case B, and the corresponding Lunduist number is around $3.3 \times 10^7$. Such a Lundquist number is already very high comparing with all the previous solar corona magnetic reconnection simulations, even though it is still lower than the Lundquist number in real solar corona ($\gtrsim 10^{12}$). The dynamic viscosity coefficient is assumed as a constant, $\nu=10^{-5}$kg\,m$^{-1}$\,s$^{-1}$. The initial $\rho_0 \eta_0$ is around $ \simeq 10^{-4}$ kg\,m$^{-1}$s$^{-1}$.

The computations are performed by using the MHD code NIRVANA \cite[version 3.6;][]{2011JCoPh.230.1035Z}. Adaptive mesh refinement (AMR) is applied. The derivatives-based mesh refinement criterion is used for the mesh refinement. The gradient-based/second-derivatives-based criterion is given by:
\begin{displaymath}
 \left[ \alpha \frac{||\delta U||_2}{|U|+U_{ref}}+(1-\alpha)\frac{||\delta^2U||_2}{||\delta U||_2+\textrm{FIL}\cdot(|U|+U_{ref})}\right]
  \left(\frac{\delta x^{(\ell)}}{\delta x^{(0)}} \right)^{\xi} \left\{ \begin{array} {ll}
     >  \varepsilon_U  \qquad \exists U   & \textrm{ refinement} \\
       < 0.8\varepsilon _U  \quad \forall U & \textrm{derefinement}
    \end{array} \right.
\end{displaymath}
where $U$ is a set of primary variables that the criterion is applied to. The mass density, momentum densities, energy density and magnetic field can all be chosen as the variable $U$ to set the criterion in the NIRVANA code. Undivided first ($\delta U$) and second ($\delta^2 U$) differences of $U$ are computed is some sort of 2-norm.
 $\alpha \in [0,1]$ (switch between a purely gradient-based criterion when $\alpha=1$ and second-derivatives-based criterion when $\alpha=0$), $U_{ref}$ (reference values),  
$\varepsilon_U$ (thresholds, the typical range is $[0.1,0.5]$) and $\xi$ (level dependence) are user-controllable parameters. The criterion is checked on a generic block octant-wise including a 2-cell wide buffer zone around the octant. FIL(preset to $10^{-2}$) is a filter to suppress refinement at small-scale wiggles. For the results presented in this manuscript, we choose the magnetic field to set the criterion. The threshold parameter $\varepsilon_U$ for the magnetic field is set to equal $0.38$, the reference value  $U_{ref}$ is $2\times10^{-5}$~T, $\alpha=0.6$. 

The time integrator for MHD equations we have used in this code is the third-order accurate Runge-Kutta method. The Second-order version of the Central-Upwind scheme is applied to the Euler equations with Lorentz-force term combined with a CT scheme for the induction equation. The electric field is computed from a genuinely 2D central upwind procedure (CCT) based on the evolution-projection method. The divergence-free condition of the magnetic field is built-in property of the scheme by virtue of a constrained-transport ansatz for the induction function. The relative divergence of the magnetic field which has been tested is normally smaller than $10^{-6}$. The detailed descriptions of this kind of scheme are presented in the paper by \cite{2011JCoPh.230.1035Z}. In that paper, numerical experiments illustrate the overall robustness and performance of the scheme for some tests. 

We start our simulations from a base-level grid of $160 \times 320$. The highest refinement level is 13, which corresponds to a grid resolution $\Delta x \approx 76$~m. It is around one magnitude higher than the ion inertial length in solar corona. Convergence studies have been carried out by repeating some simulations with a higher resolution for case B, with the highest refinement level limited to 14. The numerical results in the higher resolution case are very similar to the results presented in the next section. As we all know, the numerical diffusion is inevitable in numerical experiments. To evaluate the numerical noise in our simulations, we use the similar method as in the paper by \cite{2011ApJ...737...14S}  to perform an estimate for case B as below. The magnetic induction equation (4) can be wrote as:
  \begin{eqnarray}
    \partial_t \psi = (\textbf{v} \times \textbf{B})_z-\eta (\nabla \times \textbf {B})_z
  \end{eqnarray}   
Here the flux function $\psi$ is defined through the relations $B_x = - \partial \psi/ \partial y$, $B_y = \partial \psi/\partial x$. In the absence of numerical diffusion, both sides of Equation(16) should ideally balance each other. However, the two sides can not exactly balance each other in realistic numerical simulations. We have estimated the numerical diffusivity around the main X-point within a short time for case~B by using the following method. The main reconnection $X$-point is determined as the $X$-point which has the highest $\psi$ value of all $X$-points in the box. Suppose that the main X-point at $t=113.85$~s is at position $(x_1, y_1)$.  We define $a= \partial_t \psi(x_1, y_1)$,  $b=[\textbf{v}(x_1,y_1) \times \textbf{B}(x_1,y_1)]_z$, and $c=\eta(x_1,y_1) [\nabla \times \textbf {B}(x_1,y_1)]_z$. The values of $\lvert(a-b+c) \rvert/\lvert c \rvert$ and $\lvert(a-b+c) \rvert/\lvert (a-b)\rvert$ are calculated and presented in Figure.~2. Since the plasma velocity and magnetic field in $x$ and $y$ direction at the main X-point is near zero, the value of $\lvert b \rvert$ is much smaller than the value of $\lvert c \rvert$ in Figure.~2.  Therefore, $\lvert(a-b+c) \rvert/\lvert c \rvert \simeq \lvert(a+c) \rvert/\lvert c \rvert$ represents the ratio of $\eta_n/\eta$, $\lvert(a-b+c) \rvert/\lvert (a-b)\rvert \simeq \lvert(a+c) \rvert/\lvert (a)\rvert $ represents $\eta_n/(\eta_n+\eta)$. In Figure.~2, one can find that  the numerical diffusivity should be smaller than $20\%$ of the physical one within such a time interval. We only choose a short time interval in Figure 2, the reason is that the position of the main X-point varies with time. The main X-point is no longer at around position $(x_1, y_1)$ after $t=115.5$~s. Therefore, the value of $b$ gradually becomes larger than $c$ when the main X-point leaves away from position $(x_1,y_1)$. By using the same method, we can not prove that the numerical diffusivity in the regions away from the reconnection X-point is also smaller than the the physical diffusivity. Because the numerical diffusion is mostly caused by the $\textbf{v} \times \textbf{B}$ term in this situation, such kind of numerical diffusion could be larger than the physical $\eta \nabla \times \textbf{B}$ term.

\section{NUMERICAL RESULTS}\label{sec:results} 
\subsection{CURRENT SHEET DYNAMIC STRUCTURES}
 Disturbed by initial perturbations, the current sheet section between two primary islands as shown in Figure.~1 starts to develop towards a thinner and thinner Sweet-Parker like long current sheet. As the aspect ratio of the long current sheet exceeds a critical value, the current sheet is broken to multiple secondary fragments and many small islands start  to appear. These Secondary magnetic islands  start to grow bigger and move along the current sheet after they appear. Many newer and higher order islands also begin to develop. When we zoom in to small scales with higher resolutions as we did in Figure.~2 in our previous paper \cite[]{2015ApJ...799...79N},  we find that some secondary current sheet fragments are  broken to thinner filaments and the third order smaller plasmoids are formed. As we continue to zoom in to smaller scales, the thinnest current sheet width is then found around $1000$~m and  the highest order of plasmoids in our simulations is the fourth-order. Since there are multiple reconnection X-points,  some of the islands move upward and some of them move downward. The fast moving island can catch up with the slow island and the two islands close to each other moving with opposite directions will collide eventually. After the two islands collide, they can be coalesced to form one bigger island. The above phenomena can be seen clearly in Figures.~3, 5 and 7. From Figure.~3 and 4, we can also find that the main X-point moves up with time in our simulations, the red crosses in Figures~4(a), (b) and (c)  stand for the main X-point. The main reconnection X-points at $t=55.2$~s, $t=138.8$~s, $t=176.4$~s, $t=210.1$~s and $t=265.3$~s are detected separately at $y=1.499L_0$, $y=1.519L_0$, $y=1.582L_0$, $y=1.614L_0$ and $y=1.640L_0$. We use the same method as that in our previous papers \cite[]{2012PhPl...19g2902N, 2012ApJ...758...20N, 2013PhPl...20f1206N} to detect the main X-point. Figure.~3 and 5 demonstrate that the plasmoids above the main X-point eventually move out the simulation domain and those below the main X-point  collide with the primary big island at the bottom. Figure.~3 also shows that the maximum outflow velocity can reach around $1000$~kms$^{-1}$ even before secondary islands appear. This value is the same as the observed maximum outflow velocity. After secondary islands appear, the outflow velocities above and below the main X-point acutely fluctuate.

Figure.~3, 5 and 6 indicate that a termination shock is formed above the primary island. Figure~6(a) shows that the  plasma velocity along y direction starts to decrease to a value smaller than the sound speed $c_s$ at around $y=1.22L_0$ at $t=265.3$~s. The entropy S in Figure.~6(b) and the magnetic field parallel to the shock front in Figure.~6(c) suddenly jump to a much a higher value at around $y=1.22L_0$. This is exactly the behavior of a fast-mode shock. The termination shocks have also been found in the out-flow regions of the multiple reconnection X-points in the plasmoid dominated regime. These fast-mode shocks may be related to the hard X-ray non-thermal emission above the soft X-ray flare arcades (e.g. Masuda et al. 1994, Tsuneta \& Naito 1998, Krucker et al. 2010). However, the particle acceleration process by fast-mode shocks in kinetic scale   is still not clear and beyond the scope of our present MHD work. In the MHD scale, the hot plasmas generated by fast-mode shocks in solar corona are expected to be observed \cite[]{1979SoPh...64..287H, 2009PhPl...16i2901H}. 

The slow-mode shocks are usually formed at the outflow regions of a Petschek-like current sheet or behind the moving magnetic islands \cite[]{2001ApJ...551..312T}. In our simulations, a lot of slow-mode shocks also appear at the edges of the plasmoids. The fifth contour plot for $t=265.3$ in both Figure.~3 and 5 clearly indicate that there is a pair of slow-mode shocks at the edges of the primary island in the down flow region below the main-X point. The slow-mode shocks are also formed at the edges of secondary islands. The bottom right panel of Figure.~7(a) and bottom left panel of Figure.~7(b) show a up-moving magnetic island which is formed by two coalescent islands, a pair of nearly symmetric slow-mode shocks are formed in front of  such a island, another pair of disturbed slow-mode shocks are behind it. Except for such special cases, the slow-mode shocks are usually formed behind the moving magnetic islands in our simulations. The shock angle of the two nearly symmetric shocks is around $\approx3.7^{\circ}$. Figure.~7(c) presents the magnetic field and the current density along a cut in the $x$-direction at $y=1.697~L_0$. Around $x=0.5L_0$, the field component tangential to the shock, $B_{\parallel}$, decreases rapidly toward the downstream side and that the current density $J_z$ has a peak, but the component normal to the shock, $B_{\perp}$, stays nearly uniform along the cut. 
 
These shock structures can be distorted as the reconnection outflow plasma becomes turbulent and the plasmoids collide with each other. The distorted slow-mode shock fronts which are formed behind a up-moving magnetic island are presented in the right up panels of Figure.~7(a) and (b). The black arrows in the right up panel of  Figure.~7(b) represent plasma velocity, one can see that the parallel shear flows with different velocities appear around the contact surface between the reconnection outflows and the ambient plasmas. However, the Kelvin-Helmholtz instabilities with multiple vorticities do not appear because of the strong aligned magnetic fields\cite[]{1996ApJ...460..777F, 2002ApJ...580..800B}. These dynamic structures are only the secondary fragments of the current sheet which are fluctuated along the shock fronts.

Figure.~5 shows the temperature distributions at five different times for case B with $\beta=0.05$. During the reconnection process, one can see that the temperature gradually increases with time by ohmic heating inside the current sheet region, which is similar as those previous work (e.g., Ugai 1992). However, as shown in Figure.~4 and 5, the temperature distribution is not uniform at the current sheet region especially after secondary islands appear. Since the significant heating takes place at the slow shocks attached to the plasmoids, not at the X-points,  the temperature inside the plasmoids is usually higher than that at the reconnection X-points as shown in Figure.~4. The non-uniform behaviors of the temperature at the current regions in our simulations are in accordance with the observation results about the current sheet above the cusp-shaped structure in the gradual phase by \cite{2014ApJ...786...73S}.  In Figure.~7 in their paper, they also show the nonuniform temperature distribution along and vertical to the observed current sheet. They predict that the sharp change in both the temperature and the emission measure distribution curves could be the evidence of a slow mode shock produced by magnetic reconnection. 

\subsection{RECONNECTION RATE AND EFFECTIVE DIFFUSIVITY}
We have used the same method as those in our previous papers to calculate the reconnection rate in both case A and case B. The reconnection rate is computed as the rate of change of the magnetic flux accumulated between the $O$-point in the primary island at $y=L_0$ and the main reconnection $X$-point (see Ni et al. 2012; Ni et al. 2013),
 \begin{equation}
     \gamma(t)=\frac{\partial(\psi_X(t)-\psi_O(t))}{\partial t}\frac{1}{b_0 v_{A0}}.
     \label{e:rec_rate}
 \end{equation}              
where $b_0$ and $v_{A0}$ are the initial magnetic field and Alfv\'en velocity at the inflow boundary presented in section II, one can note that the values of $v_{A0}$ are different in case.~A and case.~B. The magnetic reconnection rates varying with the normalized timescale in Figure.~8(a) are very similar in the two cases. Secondary islands start to appear at around $0.87$\,$t_{A0A}$ in case A and $0.88$\,$t_{A0B}$ in case B, where $t_{A0A}\simeq173.66$~s and $t_{A0B}\simeq122.80$~s are the initial Alfv\'en time in case~A and case~B respectively. The reconnection rate can reach around $0.02$ in both of the two cases. However, as shown in Figure.~8(b),  the maximum temperature in case B can reach more than 2 times higher than that in case A. Therefore, the plasma can be heated to a higher temperature in a reconnection layer with a lower plasma density. The hottest plasma in case B with $\beta=0.05$ is around $30$~MK, and such high temperature plasmas can be observed by hard X ray telescopes. 

\cite{2007ApJ...658L.123L} have measured the effective magnetic diffusivity $\eta_{eff}= v_i l$ by observations, where $v_i$ is the reconnection inflow velocity and $l$ is the half-thickness of the current sheet. Figure.~9 presents the inflow velocity $v_x$ in the $x$-direction through the main X point at $t=55.2$~s and $t=265.3$~s in our simulations, here $v_x$ represents the inflow velocity $v_i$. In order to compare with observations, we measure the half-thickness of the current sheet $l$ from the position where $v_x$ starts to decrease to the position where $v_x=0$ as indicated in Figure.~9.  At $t=55.2$~s, the half-thickness is found to be $l=2\times10^7$~m and the inflow velocity is around $9\times10^4$~m\,s$^{-1}$. At $t=265.3$~s, $l$ decreases to $1.5\times10^7$~m and $v_i$ decrease to $3\times10^4$~m\,s$^{-1}$. The effective magnetic diffusivity is then obtained as $\eta_{eff} \simeq1.8\times10^{12}$~m$^2$\,s$^{-1}$ at $t=55.2$~s, and $\eta_{eff} \simeq 4.5\times10^{11}$~m$^2$\,s$^{-1}$ at $t=265.3$~s. These values are close to those deduced by \cite{2007ApJ...658L.123L} on the basis of observations. Though the resolution used in our simulations is much higher than that of the observational instruments. We should note that the refinement level for calculating the half-thickness of the current sheet in Figure.~9 is zero and the resolution is around $625$~km, which is close to the highest resolution of the solar space telescopes. The effective magnetic diffusivities at $t=265.3$~s have also been measured through other X-points  by using the same method as above. The measured values of $\eta_{eff}$ through these X-points are all on the order of $4.5\times10^{11}$~m$^2$\,s$^{-1}$.

\subsection{ENERGY CONVERSION IN CURRENT SHEET REGIONS MEDIATED BY PLASMOIDS}

Figure.~10 presents the time dependent energy conversion for $\beta=0.1$ and $\beta=0.05$ in a fixed region ($0.45L_0 \leq x \leq 0.55L_0$ and $L_0 \leq y \leq 2L_0$). 
Since energy fluxes through the boundaries at $x_b=0.45L_0$, $x_e=0.55L_0$, $y_b=L_0$ and $y_e=2L_0$ can always exist. The magnetic, thermal, and kinetic energy flowing into this region through these boundaries from the beginning of the simulation ($t=0$) to time $t$ is denoted as $E_{MF}(t)$, $E_{TF}(t)$, and $E_{KF}(t)$, respectively. (Note that these quantities may have negative signs if energy flows out of the region.)  The magnetic, thermal, and kinetic energy confined to this region at time $t$ is denoted as $E_{ML}(t)$, $E_{TL}(t)$, and $E_{KL}(t)$, respectively.  The initial magnetic, thermal, and kinetic energy at $t=0$ is denoted as $E_{MI}$, $E_{TI}$, and $E_{KI}$, respectively. The total radiated thermal energy from beginning to time $t$ is denoted as $E_{RAD}(t)$. In these notations, the dissipated magnetic energy in the region defined by $0.45L_0 \leq x \leq 0.55L_0$ and $L_0 \leq y \leq 2L_0$, is given by: $E_{MD}(t) = E_{MI} + E_{MF}(t) - E_{ML}(t)$.  In the same region, the generated thermal energy is $E_{TG}(t) = E_{RAD}(t)+E_{TL}(t) - E_{TF}(t)- E_{TI}$, and the generated kinetic energy is $E_{KG}(t) = E_{KL}{t}- E_{KF}(t) - E_{KI}$. The detailed calculations about the above variables are similar as those in our previous paper \cite[]{2012ApJ...758...20N}. However, there are some improvements in this work comparing with our previous paper. We have used 2.5-dimensional MHD instead of the 2-dimensional MHD equations,  open instead of periodic boundaries are used at the top and bottom, radiation cooling and heating terms are also included in this work. Therefore,  the $z$ components of the magnetic field and velocity should be included to calculate the above variables,  the energy flux $E_{MF}$, $E_{TF}$ and $E_{KF}$ at the boundaries and the radiated energy $E_{RAD}(t)$ have to be included. 

Figure.~10 shows that the corresponding generated thermal and kinetic energy behave similarly as the dissipated magnetic energy. From beginning to time $t$,  the total dissipated magnetic energy exactly equal the generated thermal energy plus kinetic energy, the errors are under $0.1\%$. The magnetic energy is mostly converted to kinetic energy before secondary islands appear, e.g.,  the generated kinetic energy is around four times higher than the generated thermal energy from beginning to $t=109$~s\,$\simeq 0.88 $\,$t_{A0B}$ in case B with $\beta=0.05$. After secondary islands appear, the generated thermal energy grows much faster with time than the kinetic energy, and there are more generated thermal energy than kinetic energy eventually. The small scale slow-mode shocks attached to the edges of the multiple magnetic islands play important roles in generating thermal energy. Figure.~5, and 7 clearly show that the highest temperature structures always appear at the shock front regions. In the previous papers by \cite[]{1990AN....311..399K, 2011ApJ...737...24B}, they also pointed out that the energy dissipation is accomplished via many concurrent small-scale events appearing in multiple sites distributed in space. In order to inspect the effects of Joule heating,  the generated thermal energy and Joule heating have been calculated in several fixed regions during a period.  As long as the slow-mode shocks are included in these regions, we find that the generated thermal energy is always much larger than the Joule heating in a fixed period. Therefore, Joule heating is not the main reason to cause the sharply increasing thermal energy. Though the generated kinetic energy increases slower than thermal energy after secondary islands appear, there is still around $40\%$ of the dissipated magnetic energy which has been converted to kinetic energy from beginning to $t=330$~s\,$\simeq 2.7$\,$t_{A0B}$ in case B. This is different from the result in our previous paper \cite[]{2012ApJ...758...20N}, after secondary islands appear, the generated kinetic energy decreases fast to a small value and around $99\%$ of the dissipated magnetic energy has been transformed to thermal energy eventually in that paper. The periodic boundary conditions applied in $y$ direction in the paper by \cite{2012ApJ...758...20N} is the main reason to cause this difference. The outflow plasmas are confined inside the simulation domain by periodic boundary conditions, they collide with the primary island at the top and slow down eventually. The outflow plasmas in this present work  can gradually escape from the top boundary because of the open boundary conditions. From Figure.~10(a), one can also see that the magnetic energy is dissipated faster and more thermal energy is generated in case B with $\beta=0.05$ than that in case A with $\beta=0.1$ during a period after secondary islands appear. The reason is that the plasma in the lower $\beta$ case can be compressed more strongly and the shock heating becomes more important in the energy conversion process.         

The termination shocks at the head of plasmoids also contribute a small part of the generated thermal energy. Part of the kinetic energy of the reconnection out flows can be converted to thermal energy at the termination shocks. In Figure.~5, one can see that the temperature increases at the termination shock above the primary island. However, the temperature of the heated plasma at the termination shock is much lower than the highest temperature at the slow-mode shock fronts. 

At the fragment current sheet regions where the slow-mode shocks are not included, the heating by dynamic viscosity has been measured to compare with Joule heating. The Joule heating can be measured as $Q_{\eta}=\eta (\nabla \times \bm{B} )^2/\mu_0=\eta \mu_0 \bm{J}^2$, and the viscous heating is measured as $Q_{\nu}= \nu (\nabla \cdot \bm{v})^2/3 $. At the small current sheet fragments,  the current density $J$ is around $0.1$~A\,m$^{-2}$, the maximum $\nabla \cdot \bm{v}$ is calculated around 10~s$^{-1}$, the magnetic diffusion $\eta$ is around $10^6 \sim 10^7$m$^2$\,s$^{-1}$ and $\nu = 10^{-5}$kg\,m$^{-1}$\,s$^{-1}$. Then, we can find that $\frac{Q_{\eta}}{Q_{\nu}}$ is around $30 \sim 300$. Therefore, the heating by dynamic viscosity can be ignored at the fragment current sheet regions in this work. 

\subsection{SPECTRUM STUDIES}

The one-dimensional energy spectra along the reconnection current sheet have been studied numerically in some papers \cite[]{2012PhPl...19g2902N, 2013PhPl...20f1206N, 2013PhPl...20g2114S, 2011ApJ...737...24B}. After secondary islands appear, the numerical results in these previous papers demonstrate that the energy spectrum does not behave as a simple power law anymore. The spectral index for both kinetic and magnetic energy varies with wave number,  it usually increases to a higher value as the wave number increases. Before studying the two-dimension spectra, we present the one-dimensional spectra along the current sheet for case B in Figure.~11. The similar method as that in our previous papers \cite[]{2012PhPl...19g2902N, 2013PhPl...20f1206N} has been used to get the spectral index for both kinetic and magnetic energy. Firstly, each component of magnetic field, velocity and density along the current sheet at $x=0.5L_0$ has to be transformed to Fourier space. Then, the magnetic energy spectrum is calculated as $EB_{ky} \equiv  (\tilde{B_x}^2(ky)+\tilde{B_y}^2(ky)+\tilde{B_z}^2(ky))/(2\mu)$ and the kinetic energy spectrum is $EV_{ky} \equiv  (\tilde{v_x}^2(ky)+\tilde{v_y}^2(ky)+\tilde{v_z}^2(ky))\tilde \rho_{ky}/2$. Finally, we fit the spectrum to a power law ($EB(ky) \sim ky^{-\alpha_1}$ and $EV(ky) \sim ky^{-\alpha_2}$) to obtain the energy spectrum index.  Figure.~11 shown that the spectra also do not  behave as a simple power law after secondary islands appear. We only fit a line to get the spectrum index $\alpha_1$ within the region $10^{-8} \leq EB_{ky} \leq 10^{-3} $  and $\alpha_2$ within the region $10^{-10} \leq EV_{ky} \leq 10^{-3}$. Therefore, the spectral index we have obtained is an average value. The spectral index is larger before secondary islands appear for both kinetic and magnetic energy spectrum. After secondary islands appear, the spectra become harder. The index for magnetic energy spectra decreases to around $1.8$ and the one for kinetic energy spectra is around $2.9$. These characteristics are very similar as those one-dimensional energy spectra presented in the previous papers\cite[]{2012PhPl...19g2902N, 2013PhPl...20f1206N, 2013PhPl...20g2114S, 2011ApJ...737...24B}. However, we should note that the values of these spectral index vary in an range, they are not precisely fixed. In the paper by \cite{2011ApJ...737...24B}, the one dimensional spectral index for the magnetic energy is $2.14$ at $t=316$. But such a index varies with time as shown in the papers by \cite{2012PhPl...19g2902N, 2013PhPl...20f1206N}, it is normally in the range $1.5< \alpha_1< 2.5$ for the magnetic energy spectrum after secondary islands appear. We have chosen the level 5 data to plot Figure.~11(b) and (c). At $t=55.2$~s,  the current sheet is smooth and no secondary islands appear, the highest refinement level is 4. Therefore, the level 4 data is used for plotting Figure.~11(a). 

For the first time, we have studied the two-dimensional energy spectra for both kinetic and magnetic energy. Firstly, the magnetic field, velocity and mass density in the region $0 \leq x \leq L_0$ and $L_0\leq y \leq 2L_0$ are transformed to the two-dimensional Fourier space. Then, we calculated the two-dimensional magnetic and kinetic energy as $EB_k \equiv (\tilde{B_x}^2(kx,ky)+\tilde{B_y}^2(kx,ky)+\tilde{B_z}^2(kx,ky))/(2\mu)$ and $EV_{k} \equiv  (\tilde{v_x}^2(kx,ky)+\tilde{v_y}^2(kx,ky)+\tilde{v_z}^2(kx,ky))\tilde\rho(kx,ky)/2$. There are $nx$ grids in $kx$ direction and $ny$ grids in $ky$ direction. $kx$ is defined as $0$, $2\pi/L_x$, $2\times2\pi/L_x$, $3\times2\pi/L_x$, ..., $(nx-1)\times2\pi/L_x$, and $ky$ is defined as $0$, $2\pi/L_y$, $2\times2\pi/L_y$, $3\times2\pi/L_y$, ..., $(ny-1)\times2\pi/L_y$. $L_x=L_y=L_0$ are the length scales we have selected in $x$ and $y$ direction, respectively.  The two-dimensional distributions of $lg(EB_k)$ for $t=55.2$~s and $t=138.8$~s are presented in Figure.~12. Since the high energy parts of the two dimensional spectra are mostly located at the positions with small $kx$ and $ky$. In order to see the distributions of the two dimensional spectrum more clearly. The imaginary parts of the spectra with negative coordinates are also presented in Figure.~12. We have also zoomed into a smaller scale within $ 0 \leq kx \leq 160\times 2\pi/L_0$ and $ 0 \leq ky \leq 160\times 2\pi/L_0$. Then, one can clearly see that the high energy parts are located in the center of the plots. In Figure.~12, we only need to focus in the first quadrant (the top right quadrant) with positive coordinates in the two panels. The dark red color represents the highest magnetic energy and the dark blue color represents the lowest one. At $t=55.2$~s before secondary islands appear, the distribution of magnetic energy in Fourier space is relatively smooth and the high magnetic energy is confined in the region near $ky=0$. At $t=265.3$~ s after secondary islands appear, part of the high energy is obviously spread in to the space with larger $kx$ and $ky$. As we know, the larger the wave number is, the smaller the wave length is. These evidences prove further that plasmoid instabilities can cause multiple levels of fine structures in the reconnection regions. The magnetic energy can be cascaded to smaller and smaller scales, and it is dissipated in these small scales eventually. Comparing with a one dimensional spectrum, the two dimensional ones show how asymmetric the spectra look like and give more detailed information. In the right panel of Figure.~12, one can clearly see that the magnetic energy is not uniformly cascaded in to the the Fourier  space after secondary islands appear, and most of the magnetic energy is still confined in the scales with small $ky$. The level 4 data have been used to derive Figure.~12. The distributions of the two-dimensional kinetic energy spectra which we do not show here is very similar as the magnetic energy spectrum shown in Figure~12. 

The plasmoid distribution function is also important for revealing the statistical properties and understanding the dynamics of these plasmoids.  Following the same method as  \cite{ 2012PhPl...19d2303L} and  \cite{2013PhPl...20g2114S} did, we  have also calculated the plasmoid distribution function $f(\psi)=-dN(\psi)/d\psi$ numerically. $N(\psi)$ is the number of plasmoids with magnetic flux larger than $\psi$. The magnetic flux of a magnetic island is calculated by $\rvert \psi_X-\psi_O \lvert$, where $\psi_O$ is the magnetic flux at the $O$-point of the magnetic island and $\psi_X$ is the magnetic flux at the nearby $X$-point.  For obtaining the plasmoid distribution function presented in Figure.~13, plasmoids appearing in different snapshots with an interval of $0.048t_A$ are accumulated during the evolution of the current sheet. Figure.~13 shows that the plasmoid distribution function behaves as a power law closer to $f(\psi) \sim \psi^{-1}$ in the intermediate $\psi$ regime. In the large $\psi$ regime, the distribution function $f(\psi)$ gradually deviates from the power law $\sim \psi^{-1}$ to a more rapid falloff. 

\section{SUMMARY AND DISCUSSION}
\label{sec:conclusions}
Studying the energy conversion and spectra of a corona current sheet are very important to understand the physical mechanisms of magnetic reconnection and to explain many observational features in flares. In this work we have simulated a 2.5-dimensional corona current sheet with more realistic physical parameters,  the temperature dependent high Lundquist number ($10^6 \sim 10^7$) has been used, both the radiation cooling and heating terms are included.  Here is a summary of the main results.

1. After the Sweet-Parker long current sheet is broken to multiple fragments, many Petschek-like fine structures with slow-mode shocks attached at the edges of  the plasmoids are formed. Unlike the classical Petschek slow-mode shocks, these shock structures are not steady and they can be distorted by the colliding plasmoids and turbulent outflows. Lots of turbulent structures appear inside the multiple plasmoids and in the down flow region. The termination shocks can also be formed above the primary magnetic island and at the head of secondary islands. These shocks play important roles in generating thermal energy in a {corona} current sheet.

2. For a numerical simulation with initial conditions $\beta=0.05$, $b_0=0.001$~T and $\rho_0=1.2\times10^{-12}$~kg m$^{-3}$, about $80\%$ of  the dissipated magnetic energy is converted to kinetic energy before secondary islands appear. After multiple slow-mode shocks appear at the edges  of the magnetic islands, the generated thermal energy increases sharply, about $60\%$ of the dissipated magnetic energy can be transformed to thermal energy eventually.

 3. The one dimensional energy spectra along the current sheet at $x=0.5L_0$ have been studied. After secondary islands appear, the average spectrum index for kinetic energy is around $2.9$  and it is around $1.8$ for the magnetic energy spectrum. These spectra do not behave as a simple power law and  the spectrum index increases with the wave number, which are similar as the previous studies \cite{2012PhPl...19g2902N, 2013PhPl...20f1206N, 2013PhPl...20g2114S, 2011ApJ...737...24B}. For the first time, we have studied the two-dimensional energy spectra of the corona current sheet. Comparing with the one dimensional spectra, two dimensional spectra intuitively show that part of the high energy is cascaded to larger $kx$ and $ky$ space after secondary islands appear. The spectra are asymmetric in the Fourier space, most of the energy is always confined in the region with small $ky$. 

4. The plasmoid distribution function has been calculated numerically by $f(\psi)=-dN(\psi)/d\psi$. It behaves as a power law closer to $f(\psi) \sim \psi^{-1}$ in the intermediate $\psi$ regime, which is the same as the result from \cite{2012PhRvL.109z5002H}. In the large $\psi$ regime, the distribution function $f(\psi)$ gradually deviates from the power law $\sim \psi^{-1}$ to a more rapid falloff.  However, the exponential tail in the large $\psi$ regime as presented in the paper by \cite{2012PhRvL.109z5002H} is not clearly identified from our data presented in Figure.~13. It is probably because that we only use the snapshots before $t=180$~s of case~B data and the plasmoids we have identified are still relative small. 

5. By using $\eta_{eff} = v_{inflow}\cdot L$, the effective magnetic diffusivity is estimated about $10^{11}\sim10^{12}$~m$^2$s$^{-1}$. It is close to the results deduced by \cite{2007ApJ...658L.123L} on the basis of observations.

As we know, the effect of heat conduction can smooth out the temperature in solar corona magnetic reconnection events. The numerical simulations in the papers by \cite{1997ApJ...474L..61Y, 2001ApJ...549.1160Y} have proved this point. Our previous paper \cite{2012ApJ...758...20N} also studied the effect of anisotropic heat conduction on the magnetic reconnection process. Since the time step $dt$ for explicit scheme used in the NIRVANA code is limited by CFL condition, $dt \leq (dx)^{2}/{2\kappa}$,  where $dx$ is the space step and $\kappa$ is the thermal conductivity coefficient. After secondary islands and smaller scale current sheet fragments appear, the extremely small $dx$ and large $\kappa$ make the time step $dt$ becoming too small to continue the simulations for including the heat conduction in both case A and case B. Therefore, the heat conduction terms are not included in this work. The characteristic time scales of heat conduction in both the directions parallel and perpendicular to magnetic fields have been analytically calculated to compare with the Alfv\'en crossing time. We find that  the heat conduction effect could be very efficient at the high temperature regions in the direction parallel to magnetic fields, but it can be ignored in the direction perpendicular to the magnetic fields. As shown in Figure.~5 and 7, much higher temperatures and steep temperature gradients are built up in the plasmoids and at the slow-mode shocks and X-points. Conductive energy transport from these regions into the inflow regions and into the big plasmoid at the bottom end of the current sheet is largely directed across the magnetic field. Therefore, even though the anisotropic heat conduction is included,  the high temperature plasmas can still be confined inside the plasmoids in 2.5-dimensional simulations. But, the heat conduction can be efficient in the third direction and the maximum temperature in plasmoids (flux ropes) will become smaller in full three dimensional space than that in 2.5-dimensional simulations.  

As mentioned in the paper by B{\'a}rta et al.(2011a, 2011b), the actual physical mechanism that provides the energy transfer from the global scales, at which the energy is accumulated, to the much smaller scales, at which the plasma-kinetic dissipation takes place, is an open issue. The spectrum studies presented in Figure. 11, 12 and 13 prove that the multiple cascading process is happening or already happened in the current sheet region, both kinetic and magnetic energy are cascaded from large scales to small scales during the plasmoid cascading process. Therefore, such kind of turbulent reconnection with multiple dynamic structures can explain well the energy transfer process in the solar flare eruption.  The HXR and radio observations (e.g., Karlick{\'y} et al. 1996, 2000) also indicate that the particle acceleration takes place via multiple concurrent small-scale events distributed turbulently in the flare volume, rather than by a single compact acceleration process hosted by a single diffusion region. Such observations are usually referred to as signatures of fragmented/Chaotic energy release in flares. The spectra presented in Figure. 11, 12 and 13 can not be directly compared with the observation spectra right now. In the future, we hope that the dynamic structures in the plasmoid instability process in our simulations can be used to  study the particle acceleration by the testing particle method (e.g., Li  \& Lin 2012). A huge number of particles can be placed separately in the dynamic structures with turbulent current sheet fragments, slow-mode shocks or fast-mode shocks. Then we can analyze how these particles will be accelerated and calculate the particle energy spectrum distributions. According to these spectra, one can expect to find out the contributions of  turbulent current sheet fragments, slow-mode shocks and fast-mode shocks to particle accelerations separately. Finally, these spectra about the particle energy and number density distributions can be compared with observations.

 \acknowledgments
We would like to thank the referee, Dr. Yokoyama Takaaki, Dr. Roney Keppens and Yi-Min Huang for their helpful comments and discussions. This research is supported by NSFC (Grant No. 11203069), the key Laboratory of Solar Activity grant KLSA201404, the Western Light of Chinese Academy of Sciences 2014, the Program 973 grants 2013CBA01503, NSFC grant 11273055, NSFC grant 11333007, CAS grant KJCX2-EW-T07 and CAS grant XDB09040202. We have used the NIRVANA code v3.6 developed by Udo Ziegler at the Leibniz-Institut f\"ur Astrophysik Potsdam. The authors gratefully acknowledge the computing time granted by the Yunnan Astronomical Observatories, and provided on the facilities at the Yunnan Astronomical Observatories Supercomputing Platform, as well as the help from all faculties of the Platform. 


\clearpage

\begin{figure*}
 \centerline{\includegraphics[width=0.28\textwidth, clip=]{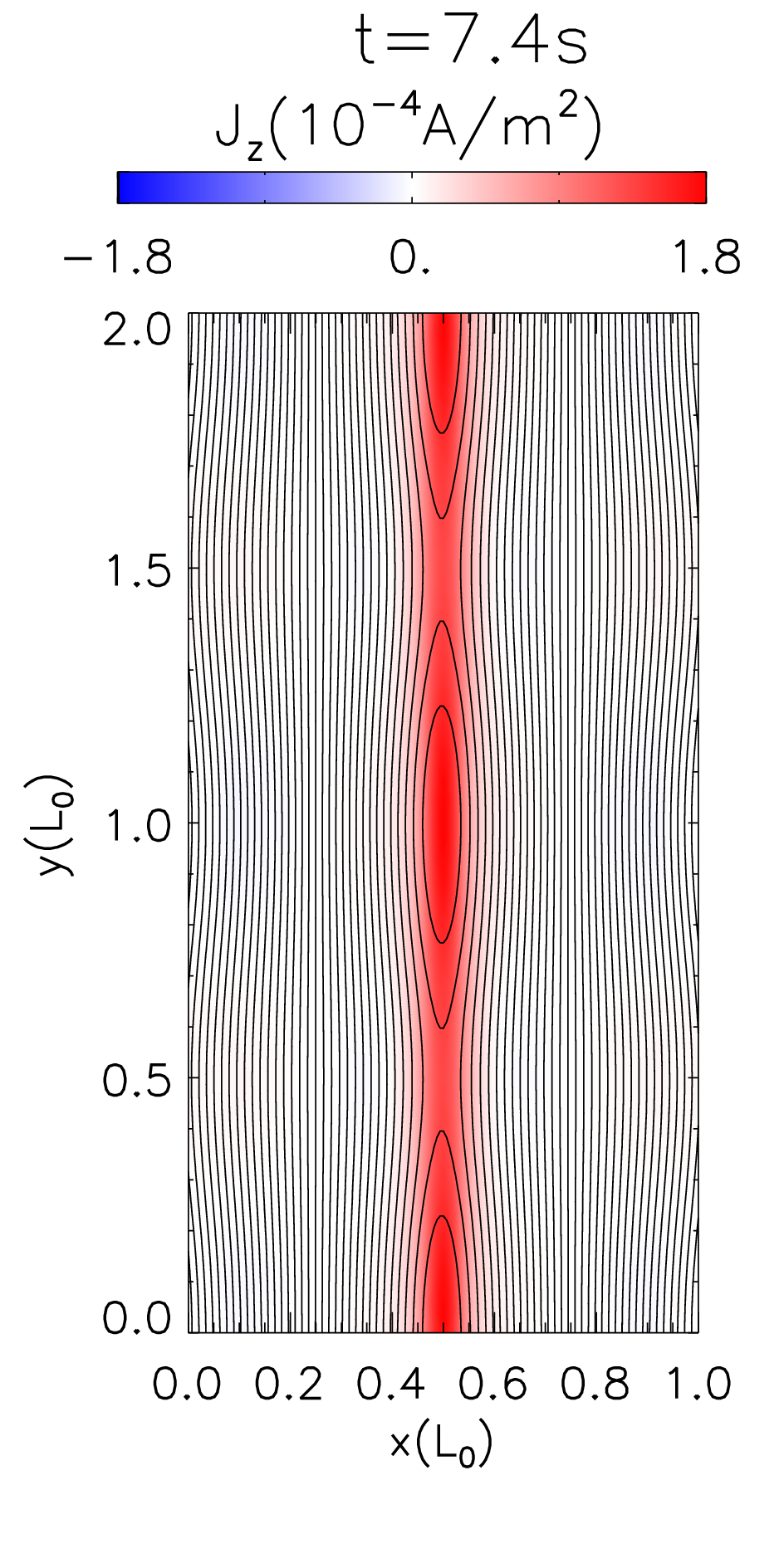}}
    \caption{Field lines and current density $J_z$ (background color) in the whole simulation domain for case~B at $t=7.4$~s.}
  \label{fig.1}
\end{figure*}

\begin{figure*}
 \centerline{\includegraphics[width=0.50\textwidth, clip=]{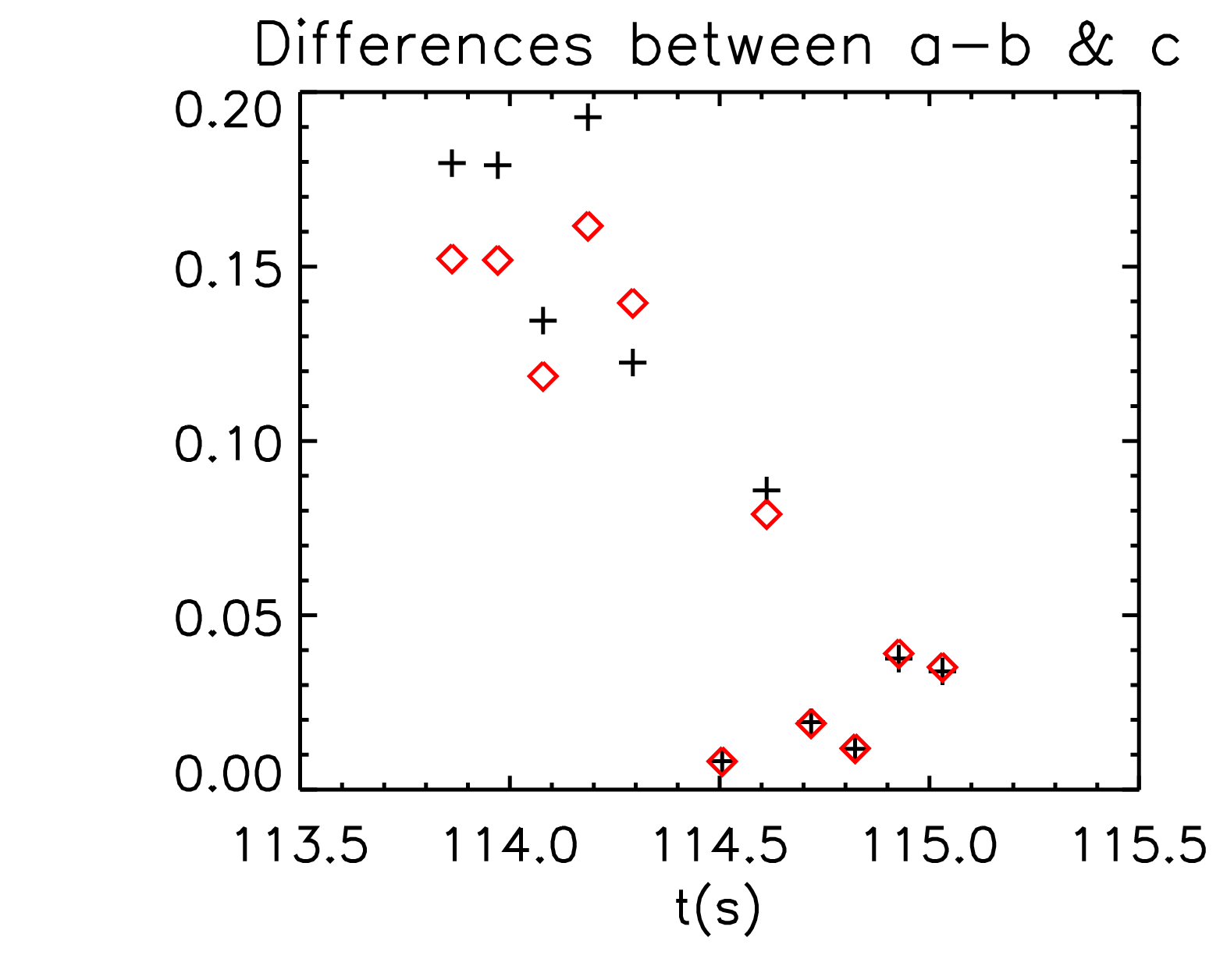}}
    \caption{The values of $\lvert(a-b+c) \rvert/\lvert c \rvert$ and $\lvert(a-b+c) \rvert/\lvert (a-b)\rvert$ at a fixed position $(x_1,y_1)$during a short time. The black crosses represent 
     $\lvert(a-b+c) \rvert/\lvert c \rvert$ and the red diamonds represent $\lvert(a-b+c) \rvert/\lvert (a-b)\rvert$. }
  \label{fig.2}
\end{figure*}

\begin{figure*}
 \centerline{\includegraphics[width=0.8\textwidth, clip=]{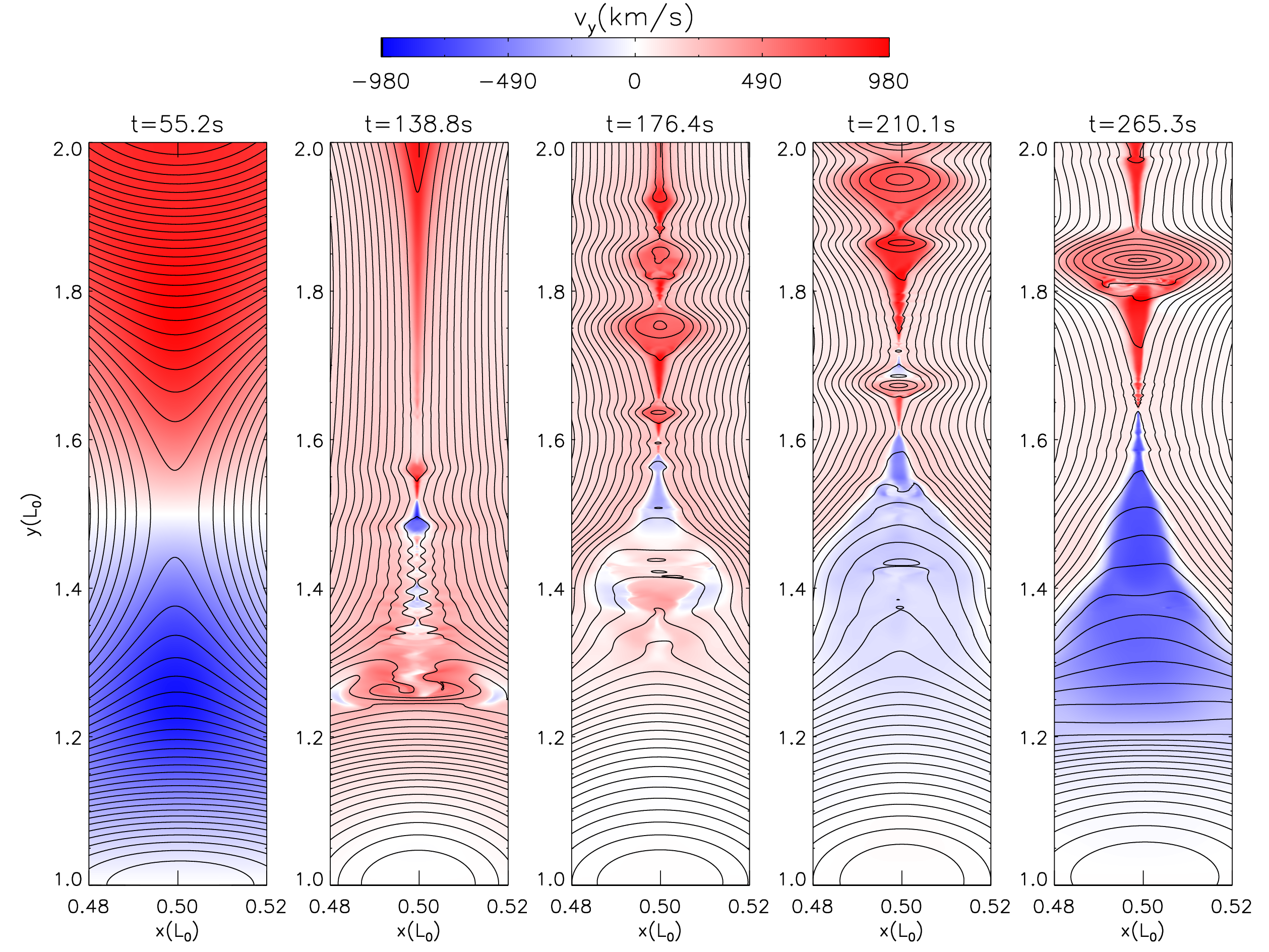}}
    \caption{Field lines and vertical velocity component $v_y$ (background color) for case~B at five different times.}
  \label{fig.3}
\end{figure*}

\begin{figure*}
  \centerline{\includegraphics[width=0.33\textwidth, clip=]{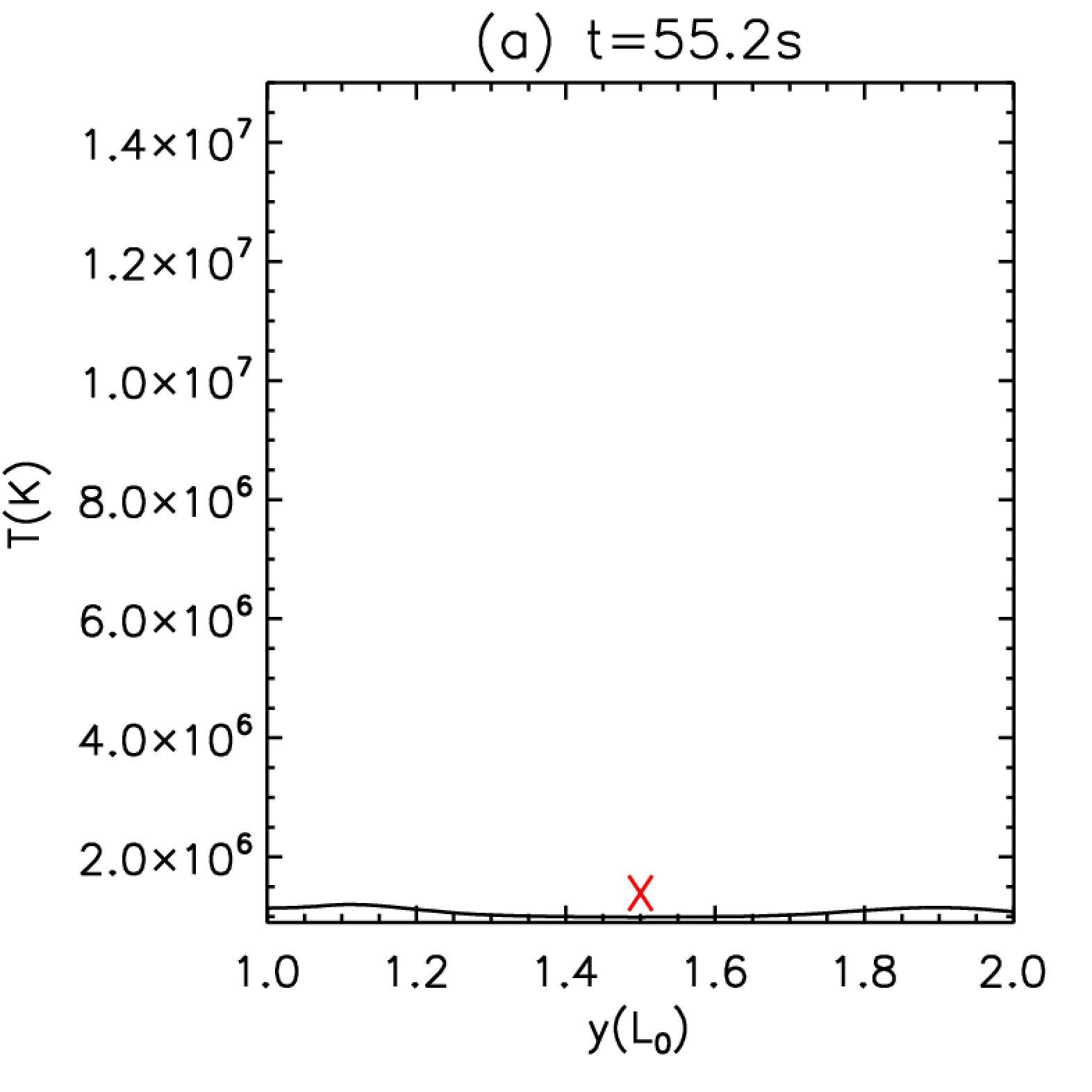}
                       \includegraphics[width=0.33\textwidth, clip=]{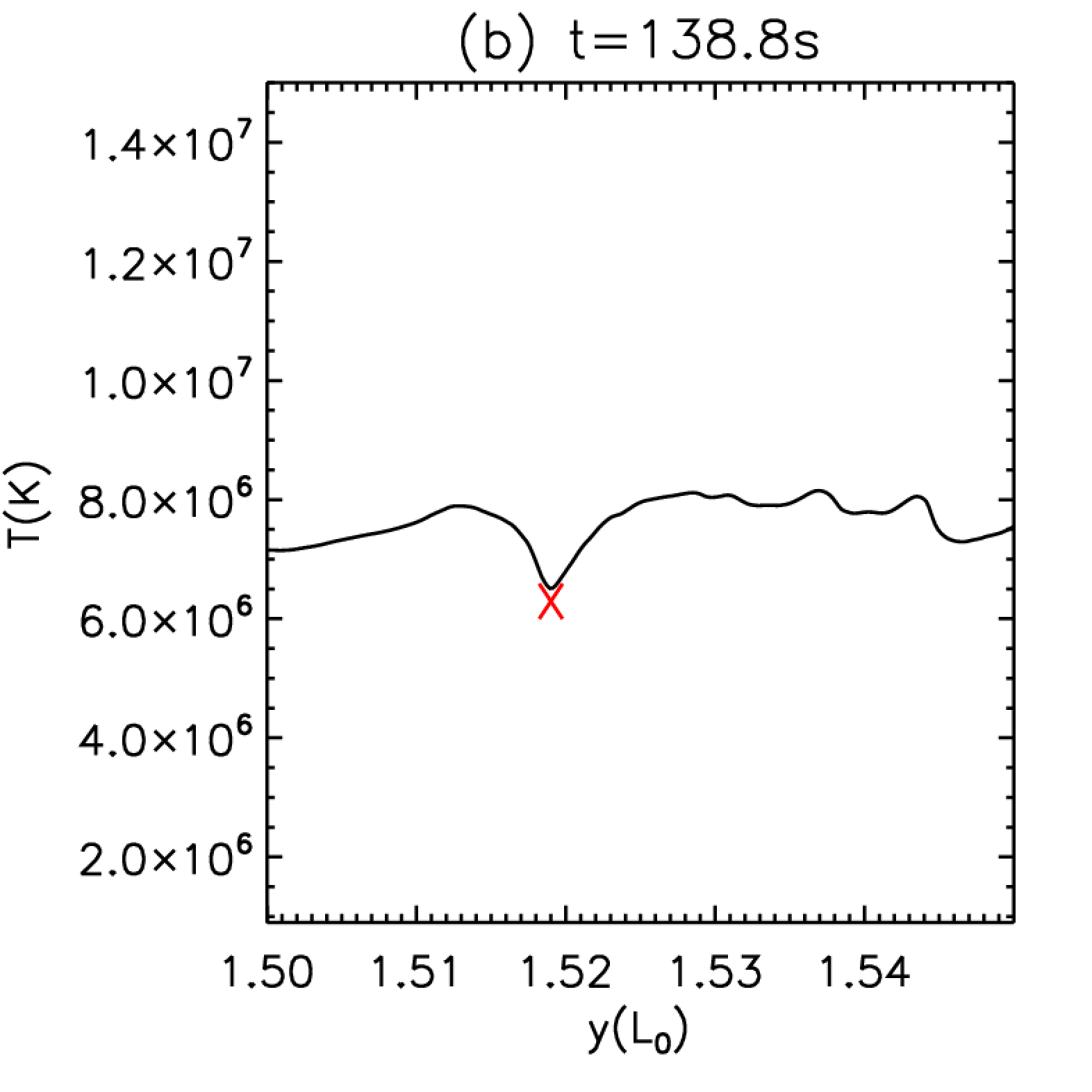}
                       \includegraphics[width=0.33\textwidth, clip=]{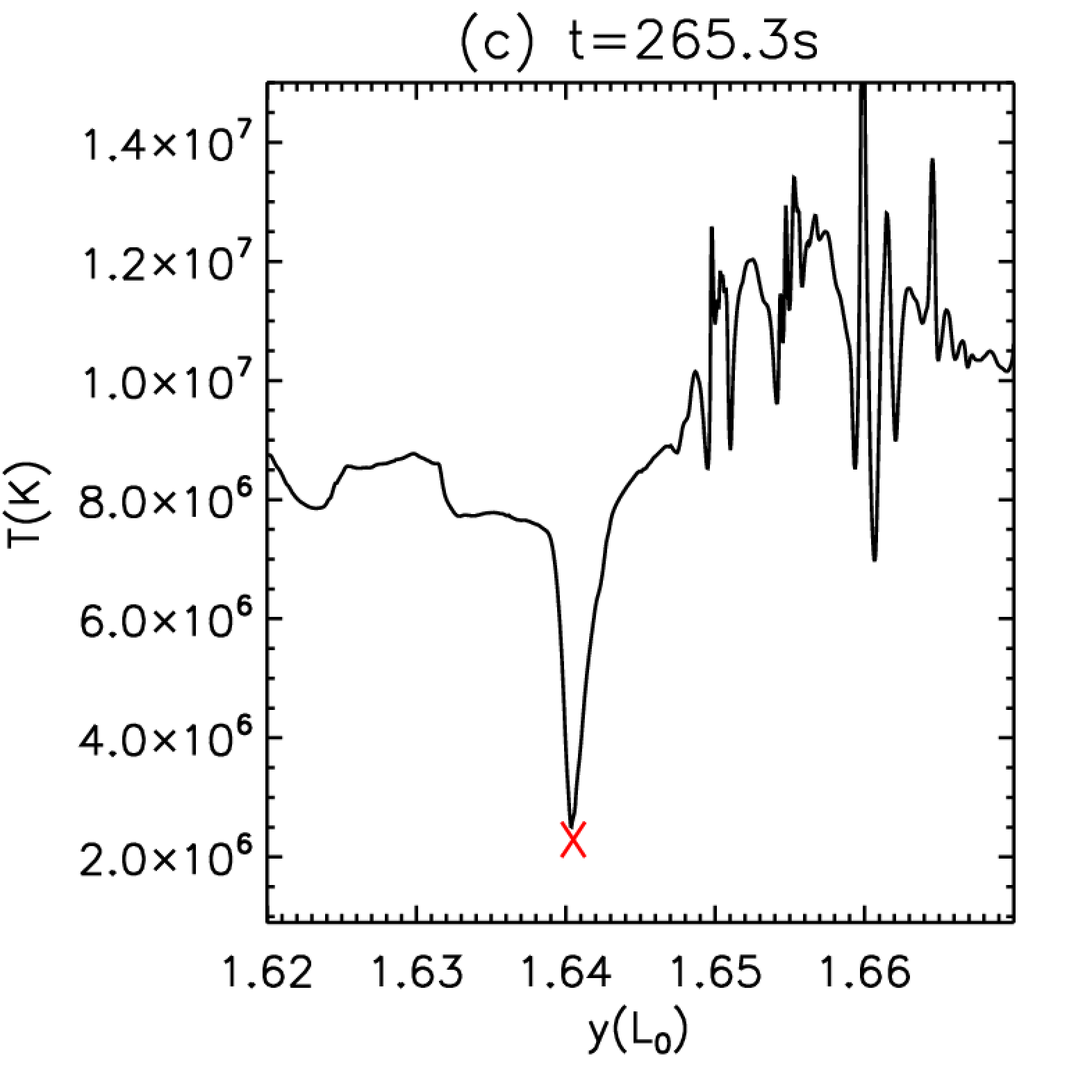}}
    \caption{The temperature of the cut at the main X-point along y direction for case~B at three different times. The red crosses in (a), (b) and (c) stand for the main-X point. }
  \label{fig.4}
\end{figure*}

\begin{figure*}
 \centerline{\includegraphics[width=0.8\textwidth, clip=]{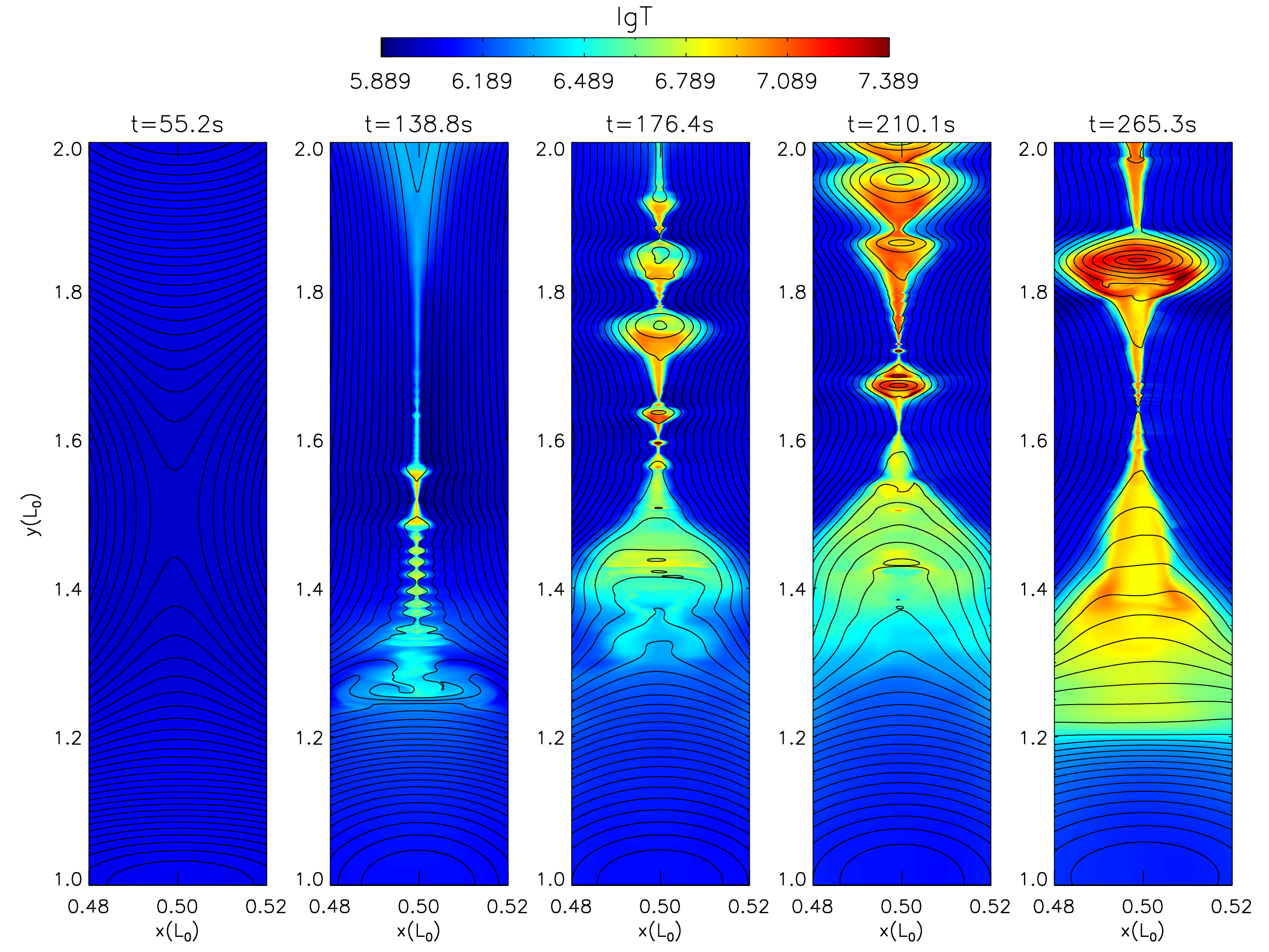}}
    \caption{Field lines and logarithmic values of temperature (background color) for case~B at five different times.}
  \label{fig.5}
\end{figure*}

\begin{figure*}
  \centerline{\includegraphics[width=0.33\textwidth, clip=]{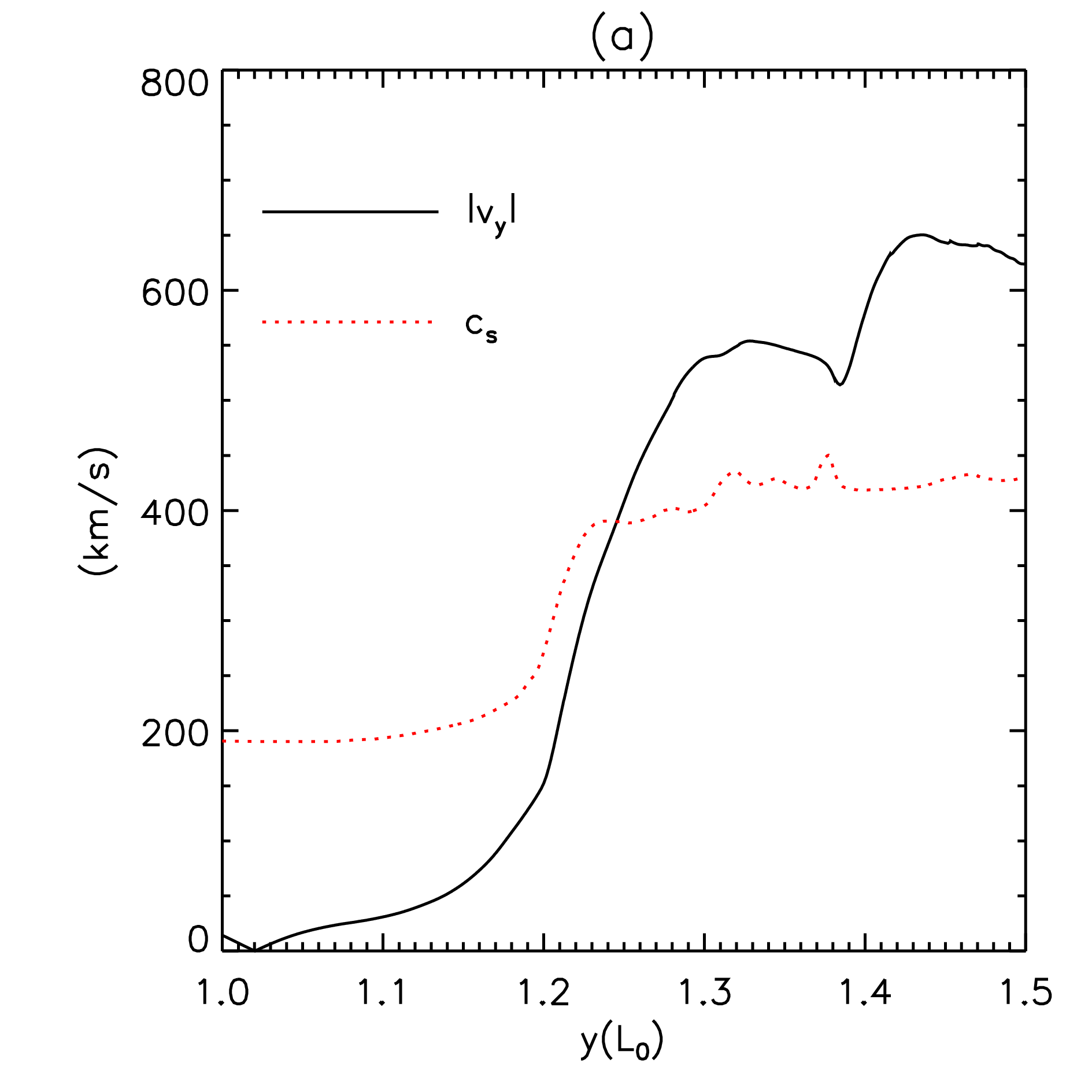}
                       \includegraphics[width=0.33\textwidth, clip=]{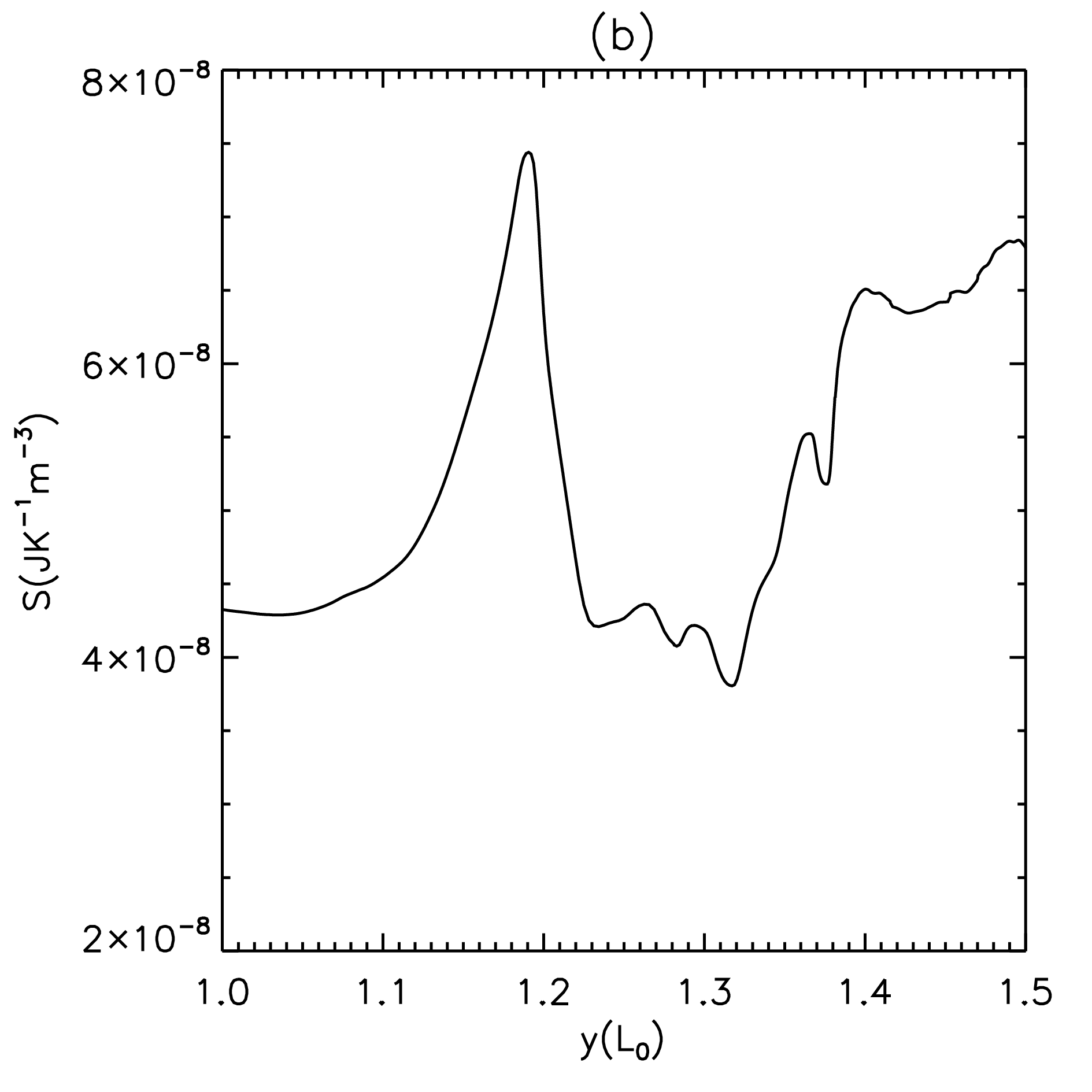}
                       \includegraphics[width=0.33\textwidth, clip=]{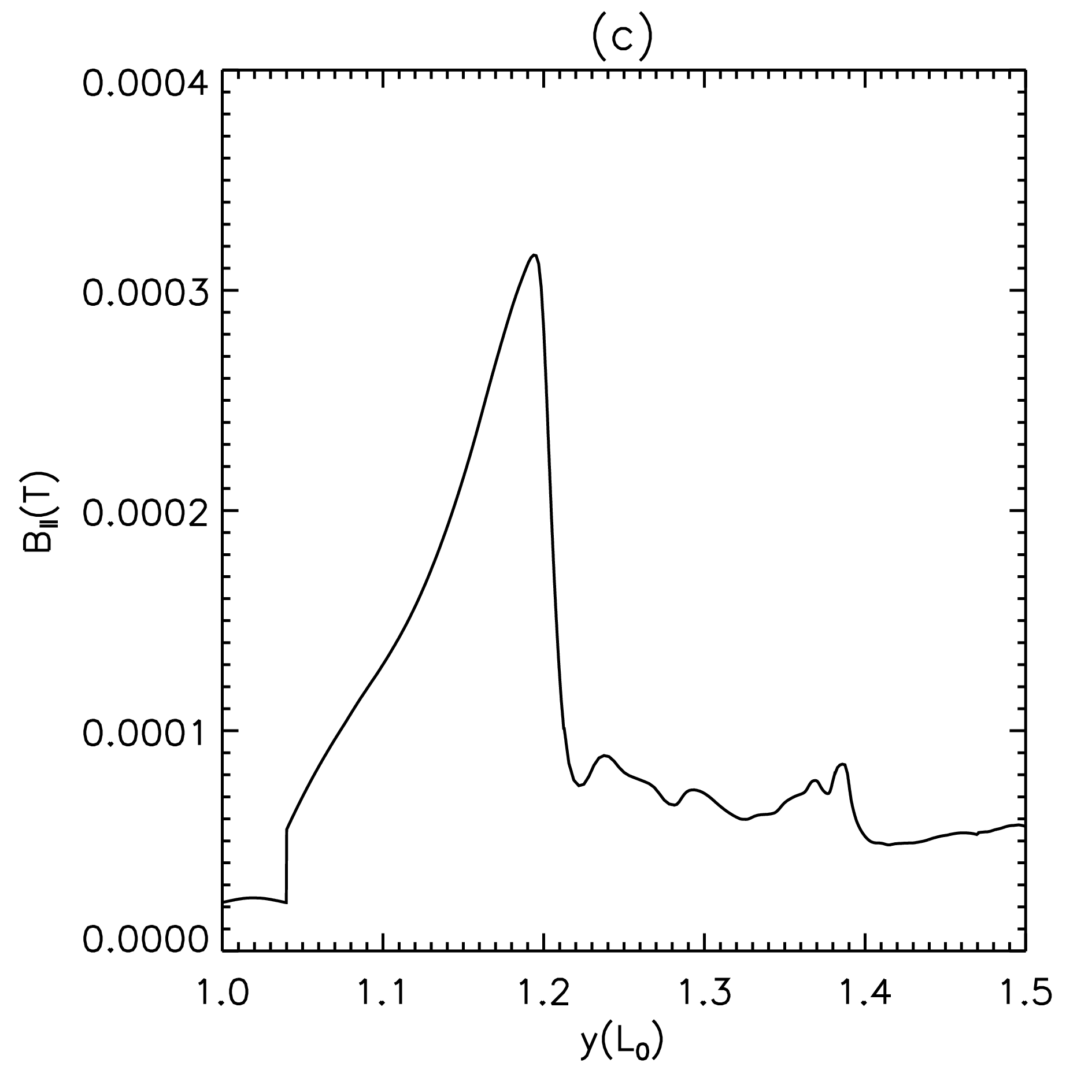}}
    \caption{At $x=0.5L_0$ along $y$-direction in the down flow region, (a) the absolute value of plasma velocity $v_y$ and the sound speed $c_s$, (b) the entropy $S$ and (c) the magnetic field parallel the shock front. }
 \label{fig.6}
\end{figure*}

\begin{figure*}
  \centerline{\includegraphics[width=0.45\textwidth, clip=]{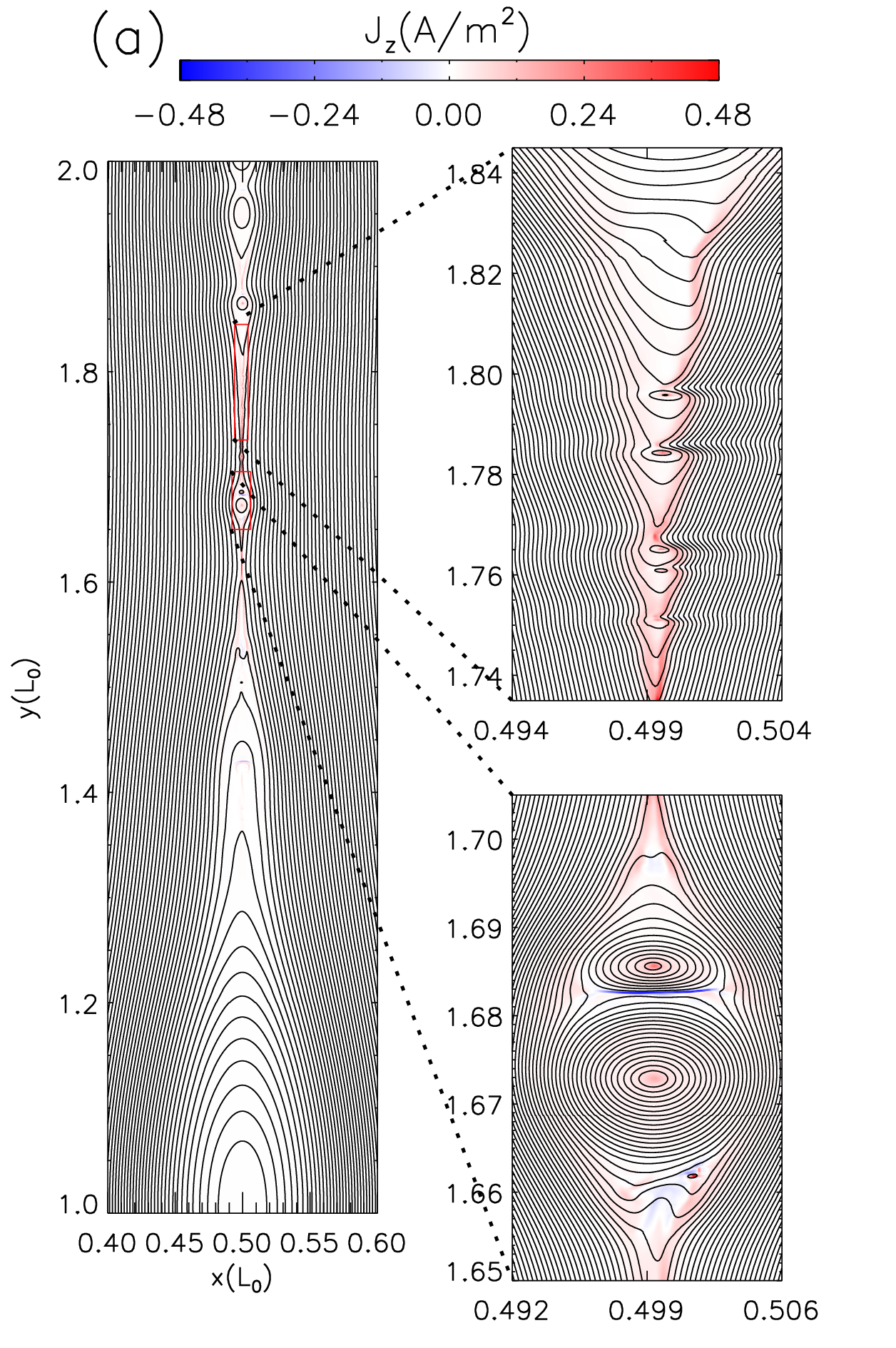}
                       \includegraphics[width=0.45\textwidth, clip=]{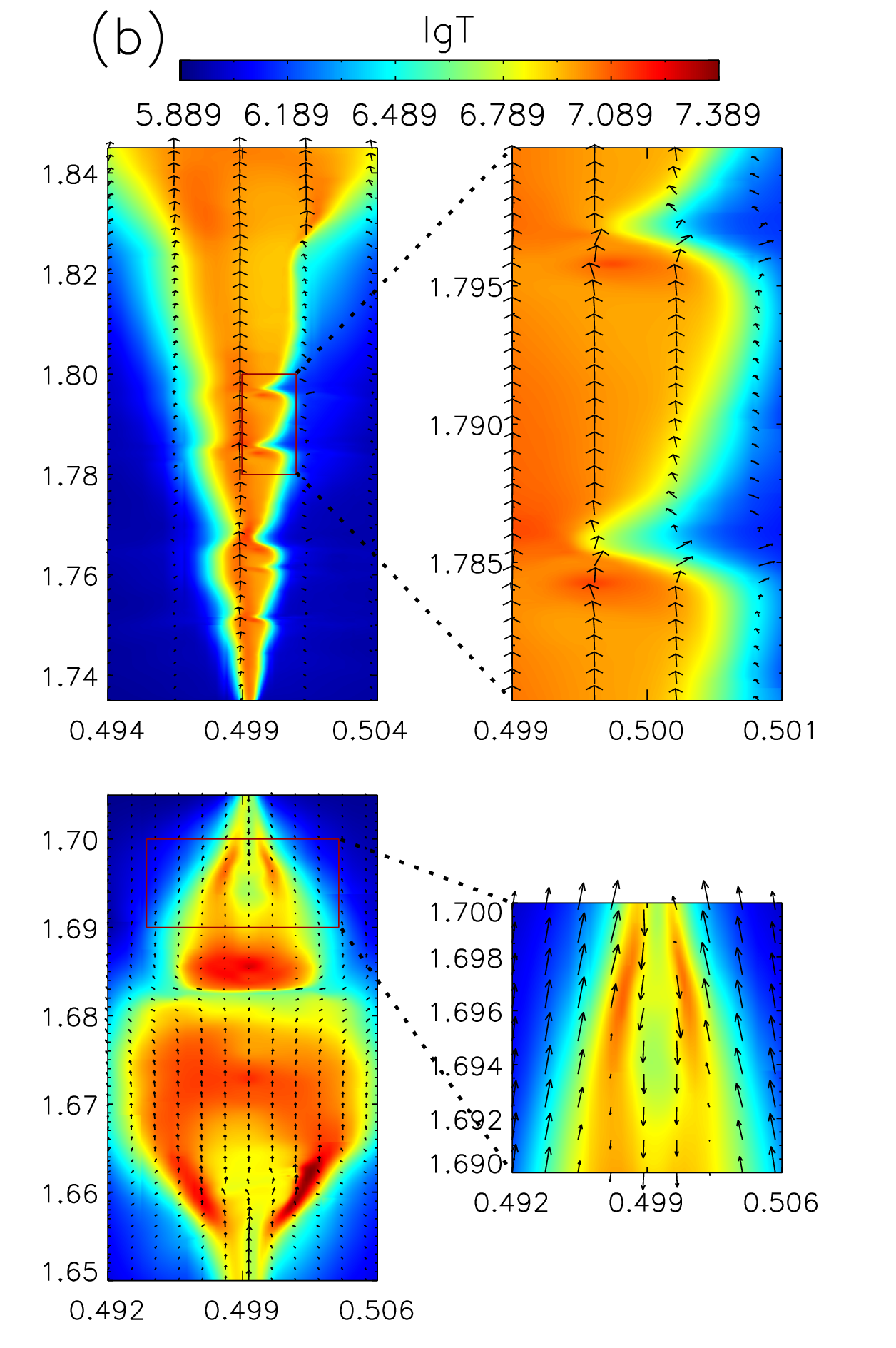}}
   \centerline{\includegraphics[width=0.45\textwidth, clip=]{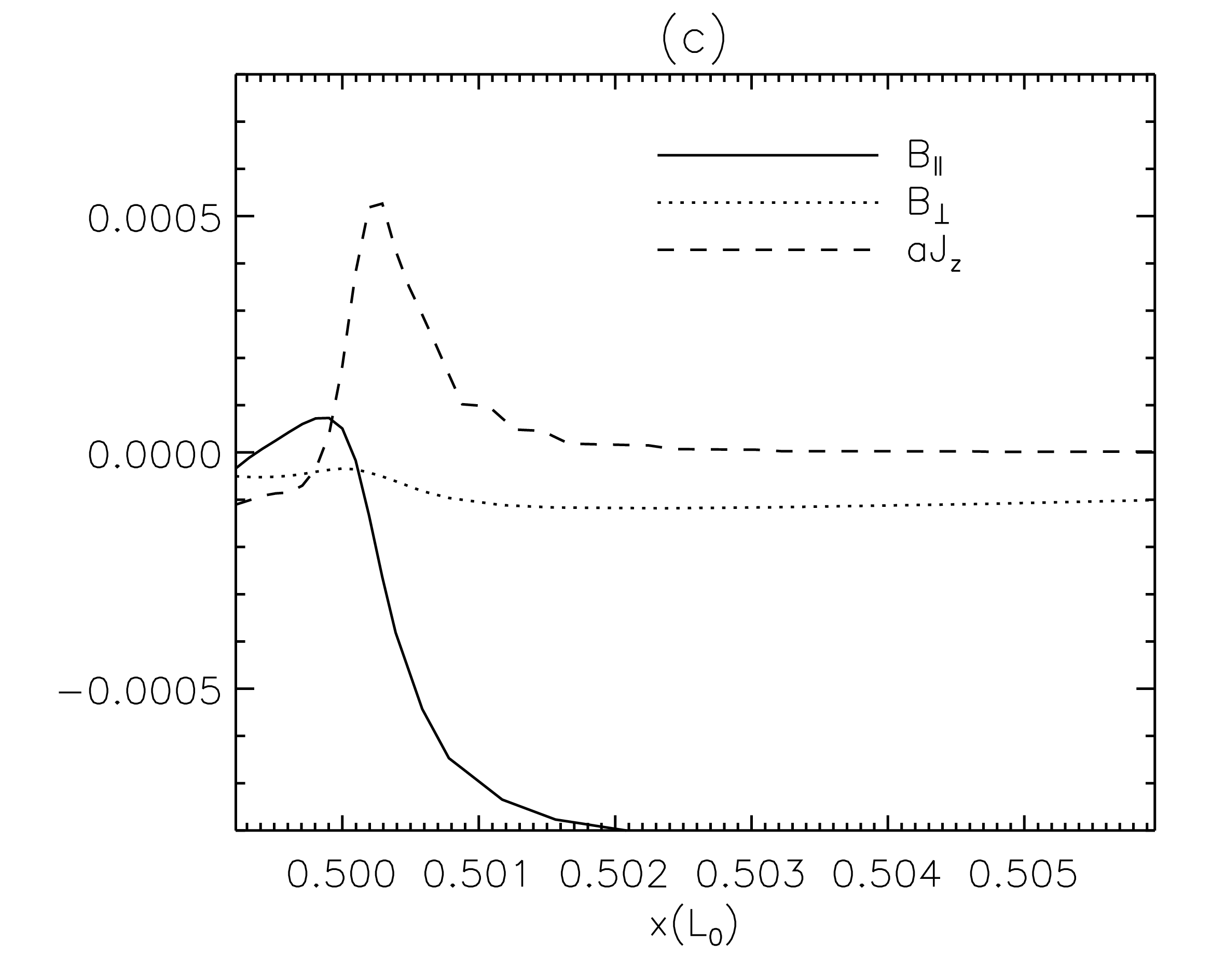}}                    
    \caption{(a)Field lines and curent density at $t=210.1$~s, two small scale regions with fragment current sheets has been zoomed in. (b) The temperature distributions and plasma velocities inside the two small scale regions, the black arrows represent the plasma velocities. (c) Magnetic field components parallel and perpendicular to the slow mode shock and current density along a cut line in $x$-direction at $y=1.697~L_0$, $a=0.005$.   }
 \label{fig.7}
\end{figure*}

\begin{figure*}
  \centerline{\includegraphics[width=0.49\textwidth, clip=]{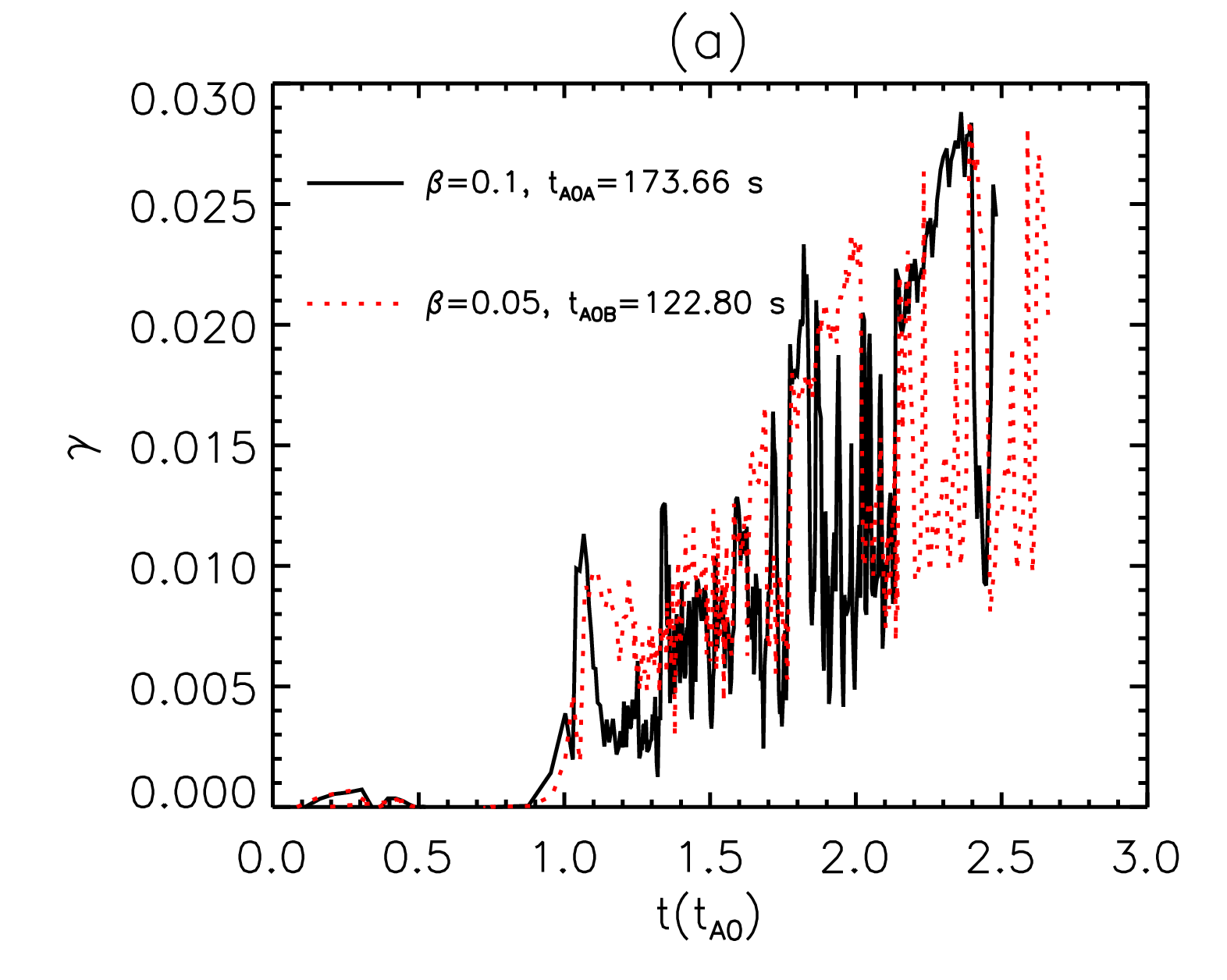}
                      \includegraphics[width=0.49\textwidth, clip=]{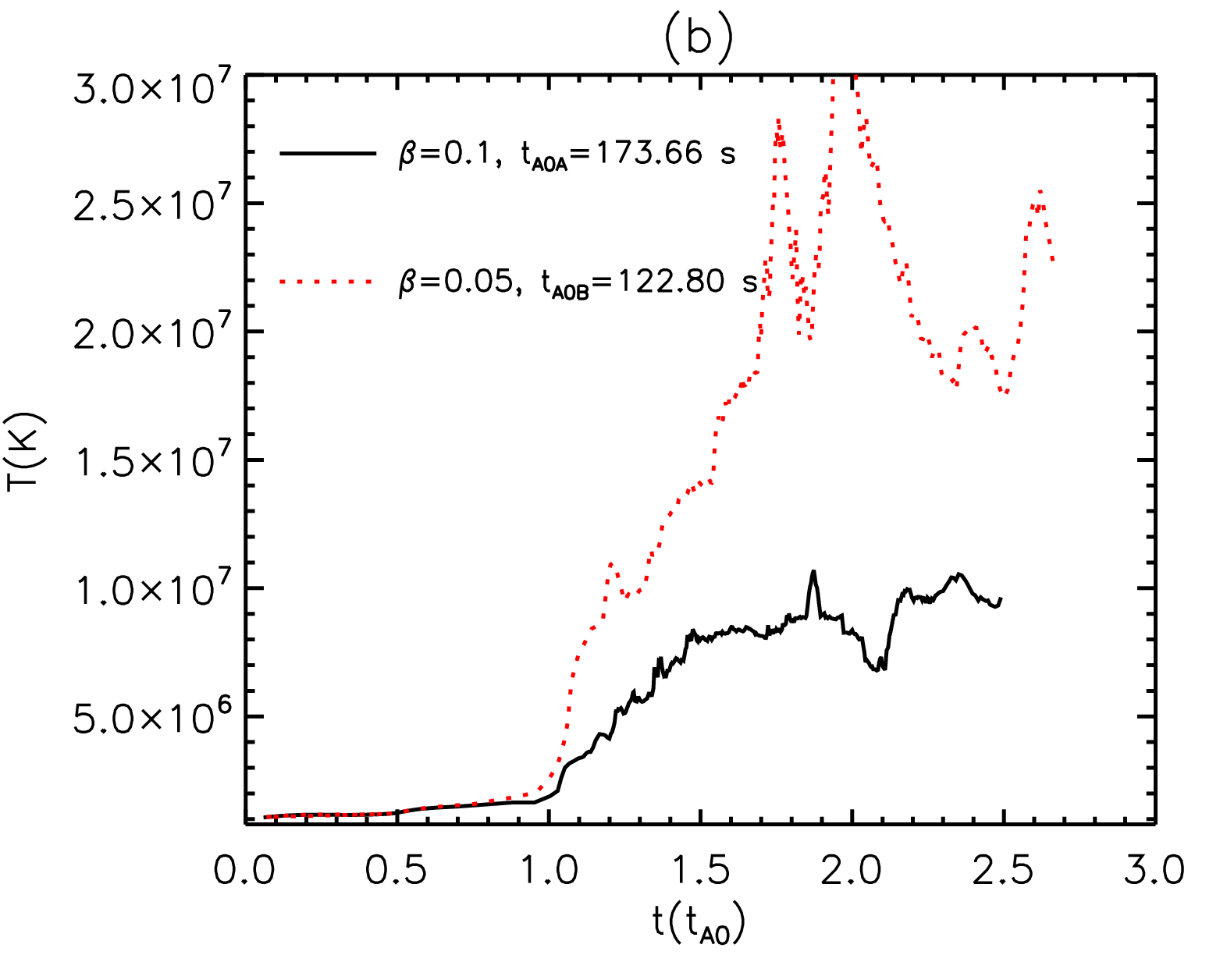}}
    \caption{(a) Magnetic reconnection rate and (b) the maximum temperature vary with the normalized time scales for case~A with $\beta=0.1$ and case~B with $\beta=0.05$}
  \label{fig.8}
\end{figure*}

\begin{figure*}
 \centerline{\includegraphics[width=0.5\textwidth, clip=]{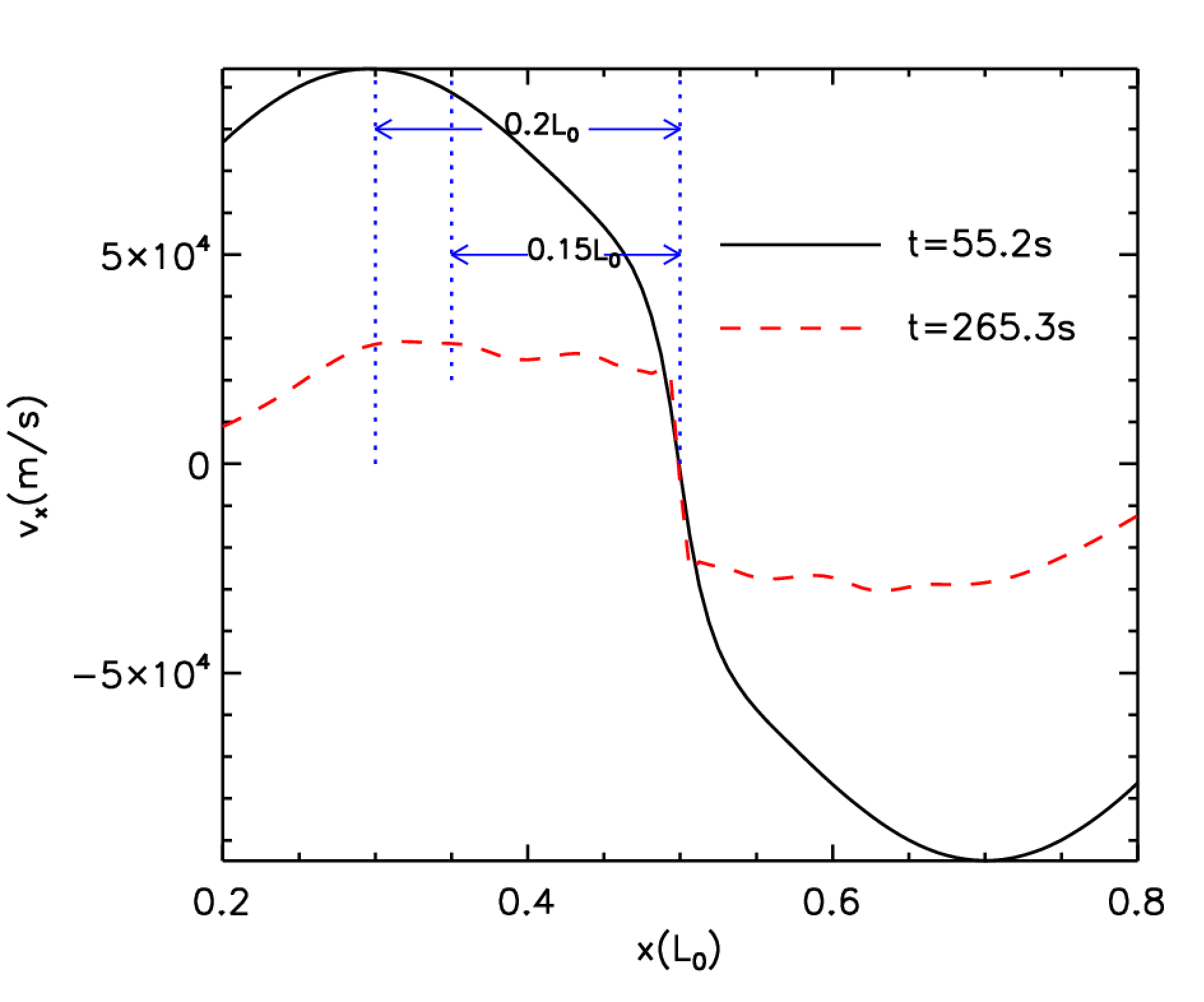}}
    \caption{The distribution of plasma velocity in x direction through the main X-point, at $t=55.2$~s and $t=265.3$~s.}
  \label{fig.9}
\end{figure*}

\begin{figure*}
  \centerline{\includegraphics[width=0.33\textwidth, clip=]{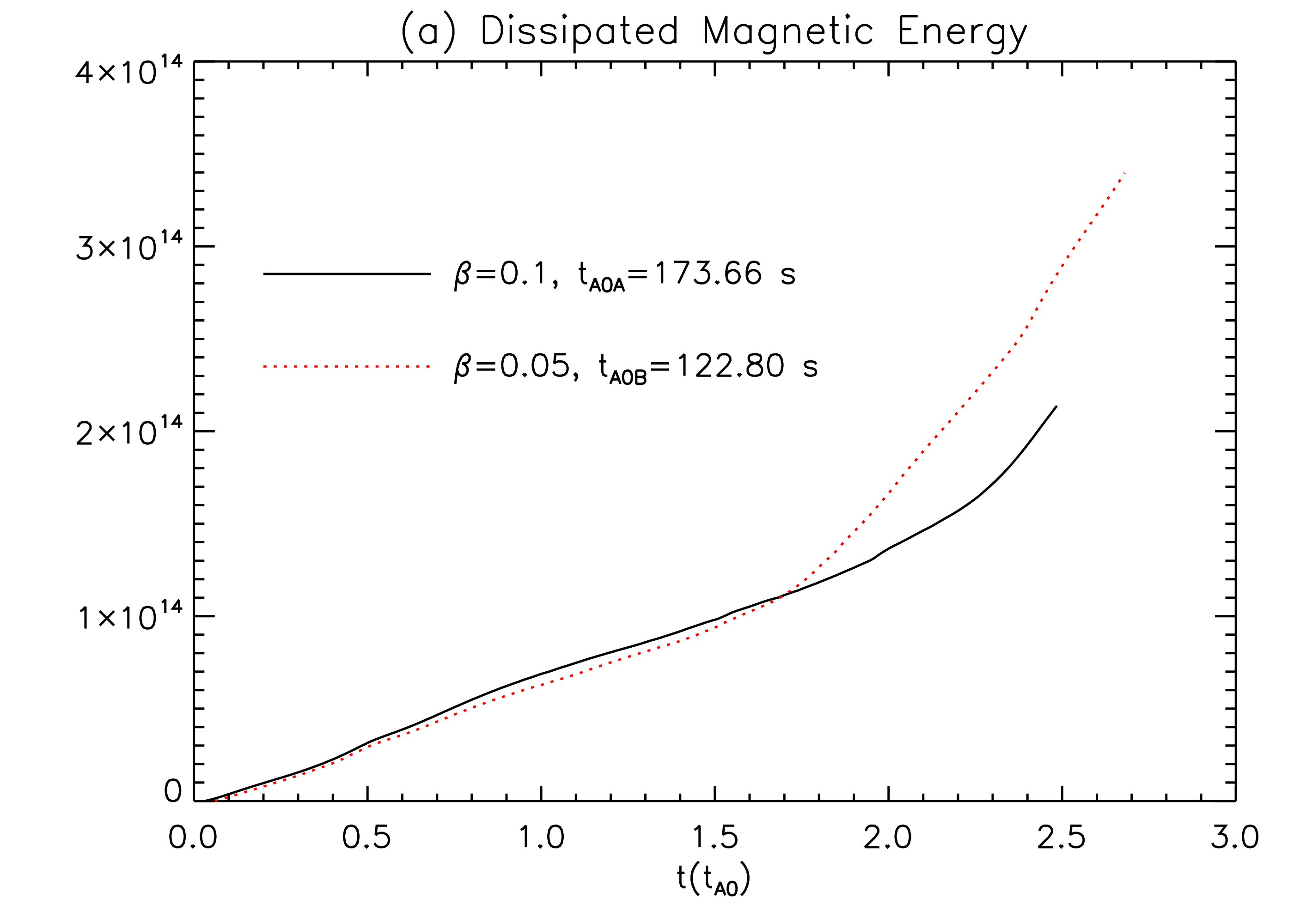}
                       \includegraphics[width=0.33\textwidth, clip=]{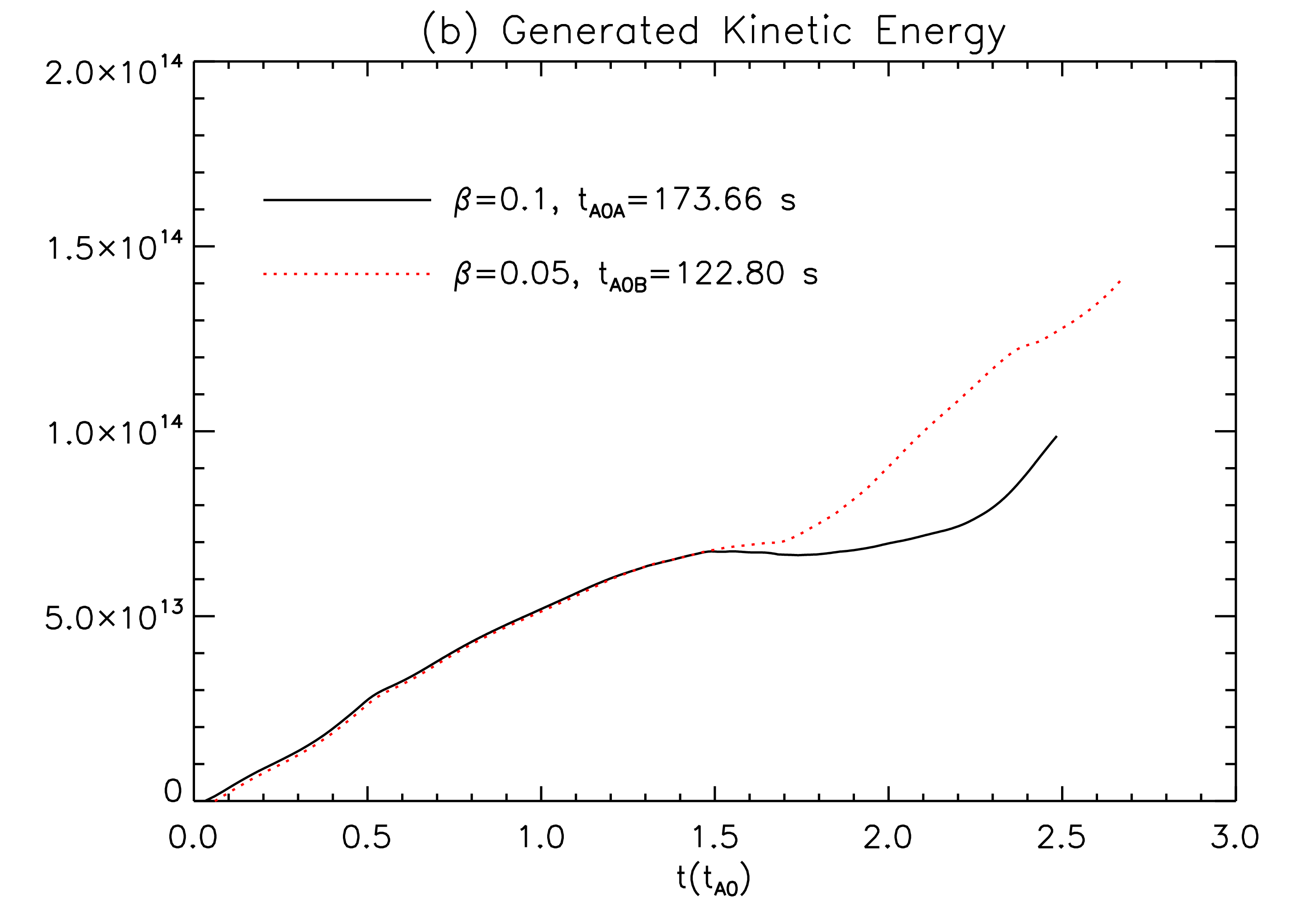}
                       \includegraphics[width=0.33\textwidth, clip=]{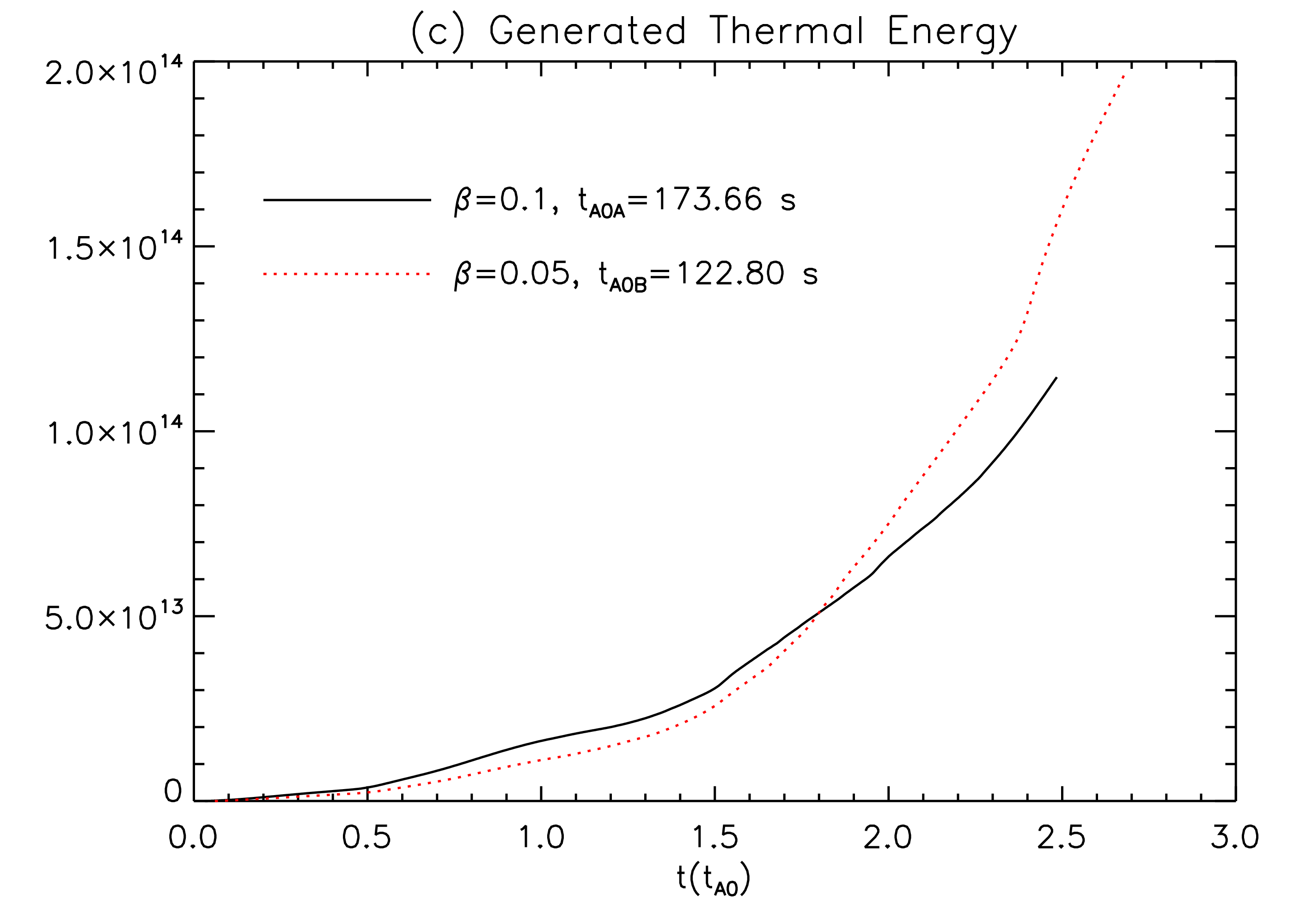}}
    \caption{The time-dependent (a) dissipated magnetic energy, (b) generated kinetic energy and (c) generated thermal energy in the dissipation domain defined in the main text.  }
  \label{fig.10}
\end{figure*}

\begin{figure*}
  \centerline{\includegraphics[width=0.33\textwidth, clip=]{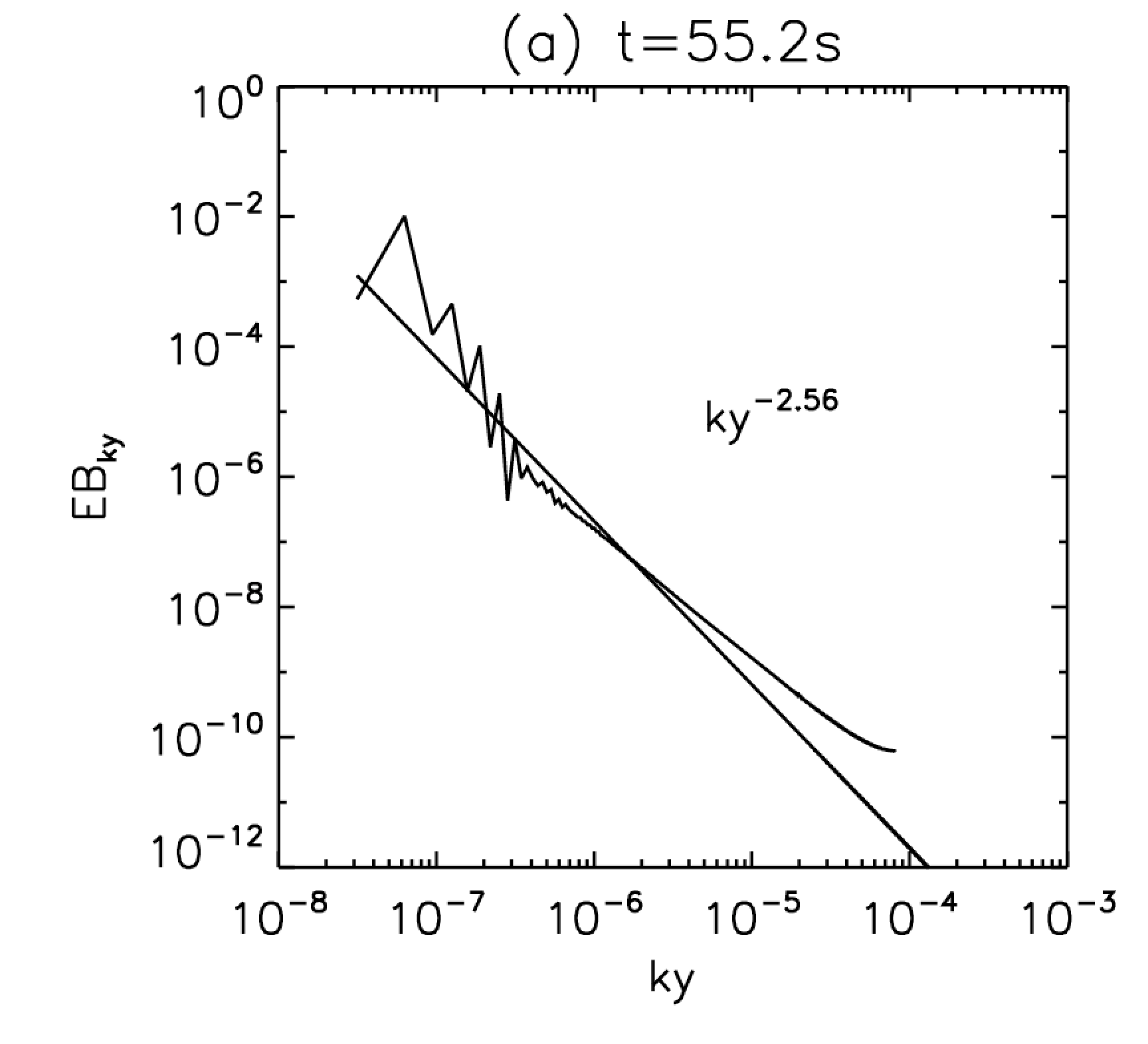}
                       \includegraphics[width=0.33\textwidth, clip=]{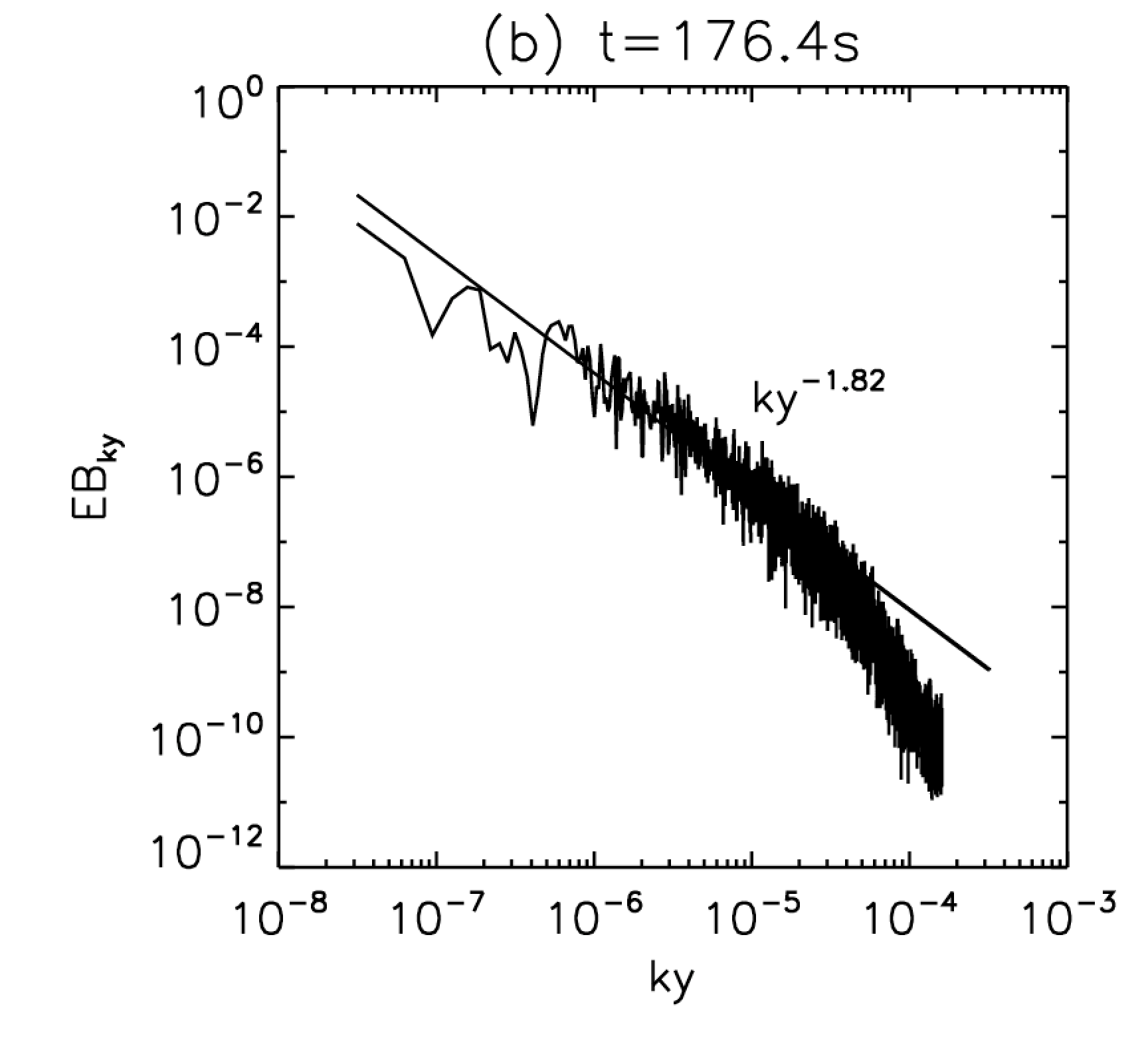}
                       \includegraphics[width=0.33\textwidth, clip=]{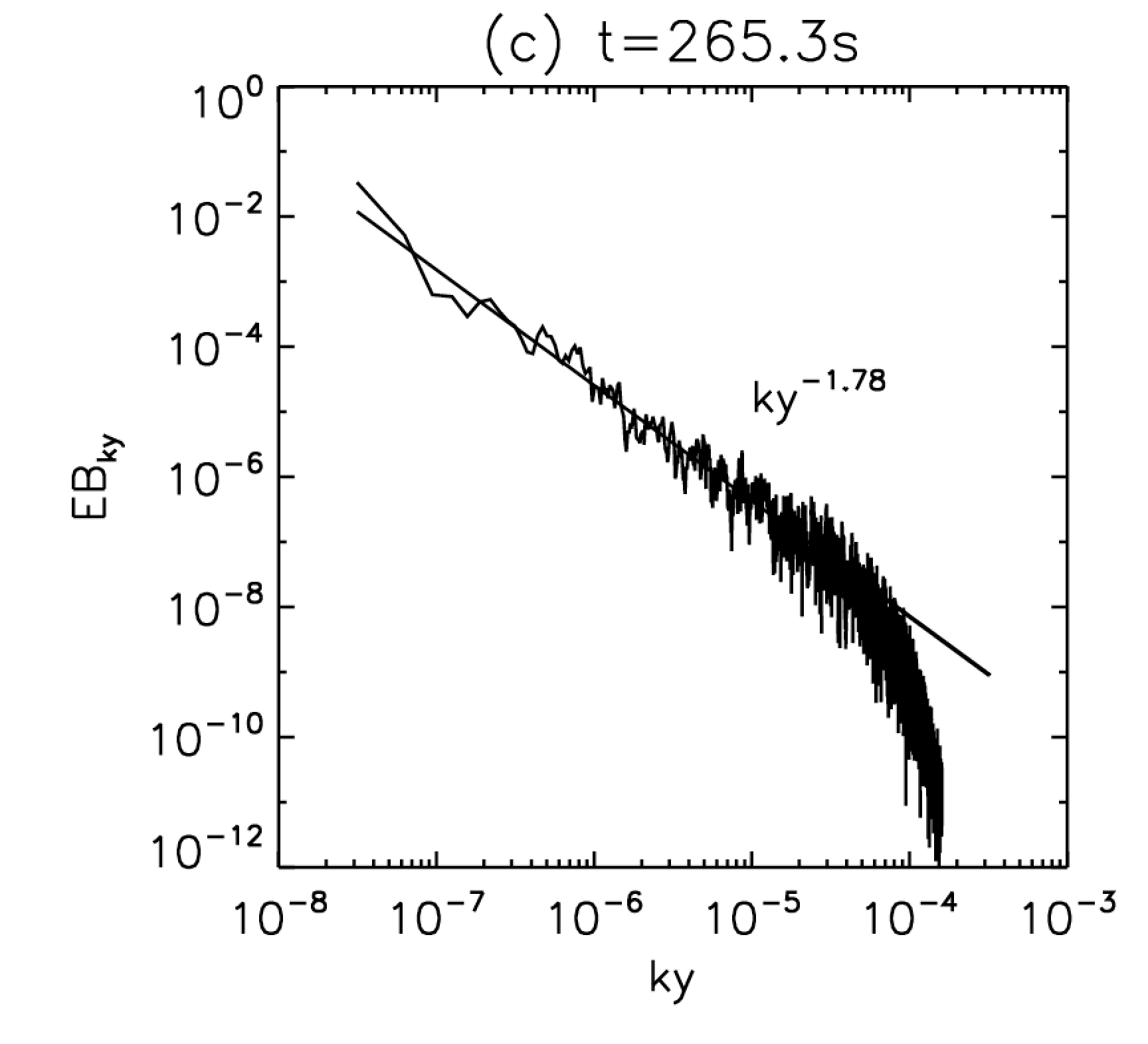}}
    \centerline{\includegraphics[width=0.33\textwidth, clip=]{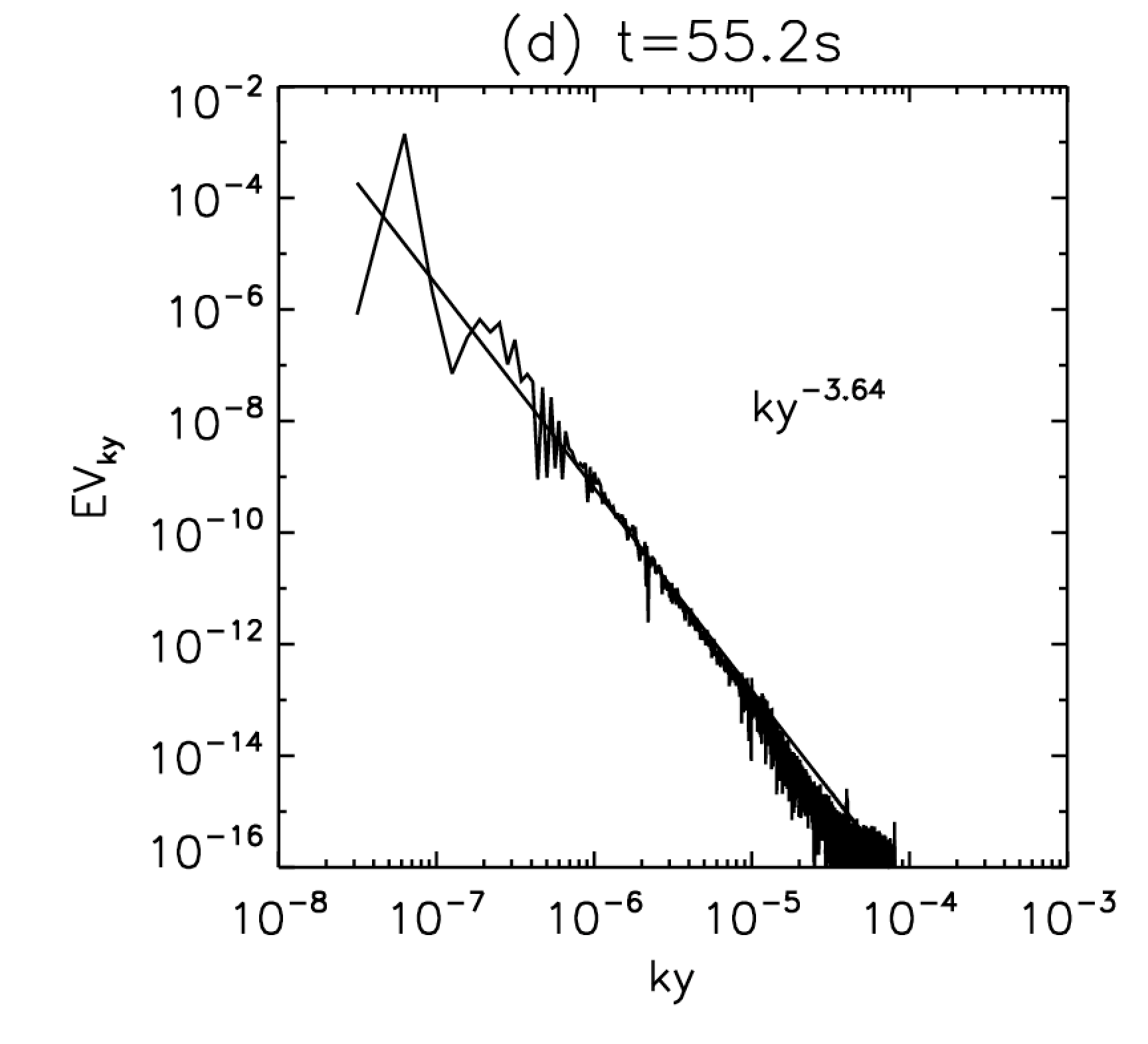}
                       \includegraphics[width=0.33\textwidth, clip=]{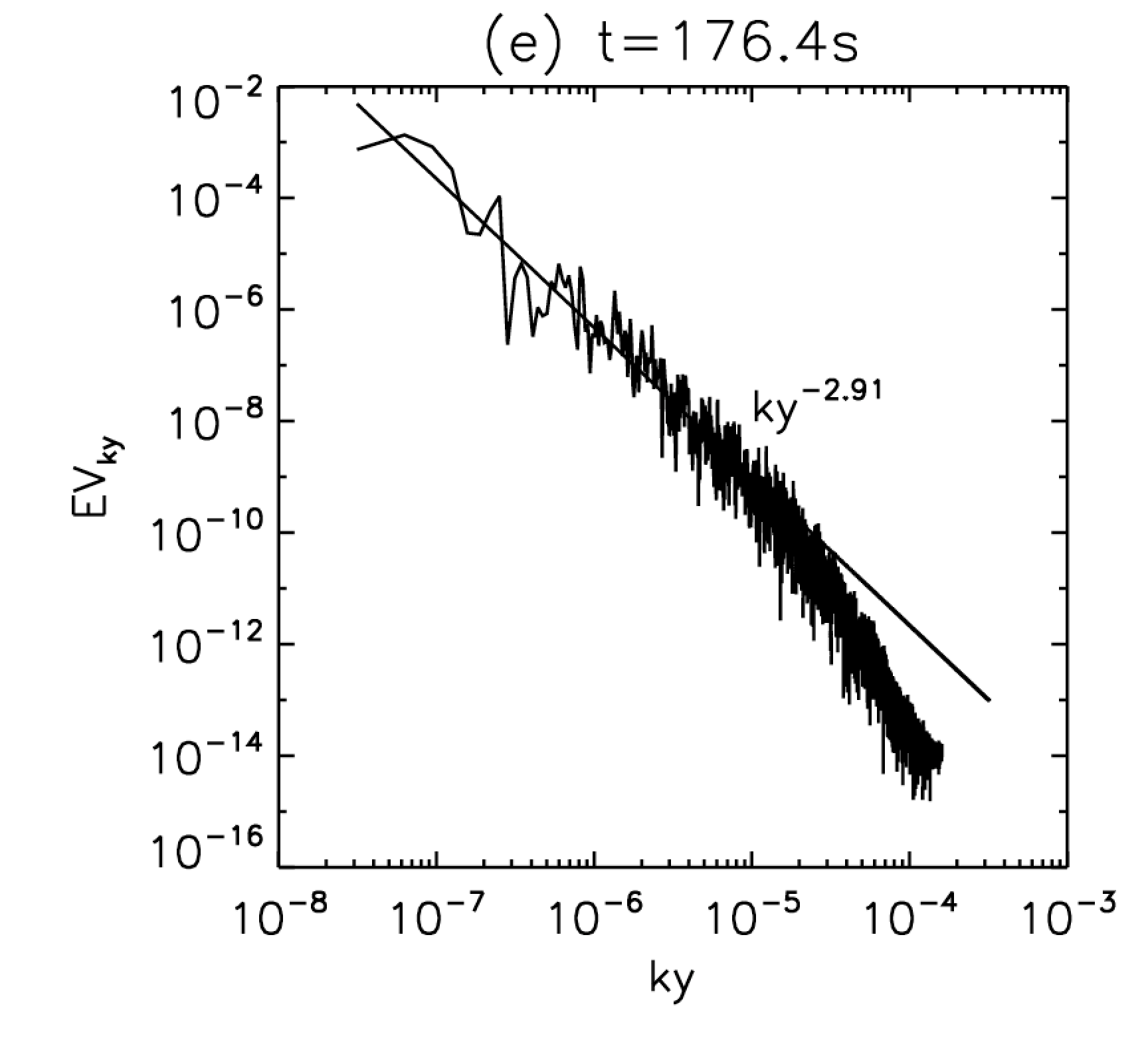}
                       \includegraphics[width=0.33\textwidth, clip=]{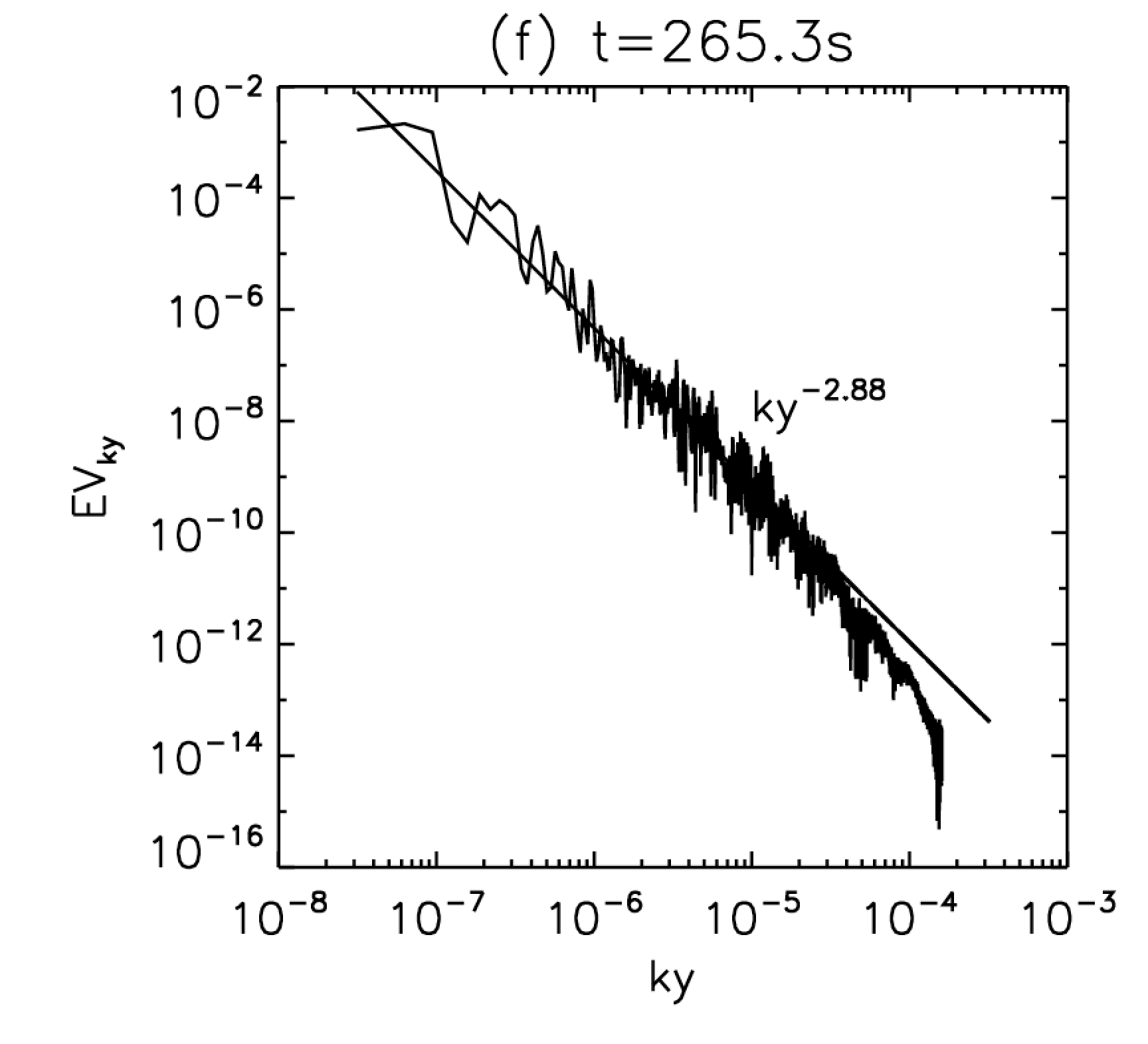}}
   \caption{The one dimensional spectrums along the current sheet at $x=0.5L_0$ at three different times; (a), (b) and (c)  are for the magnetic energy;  (d), (e) and (f) are for the kinetic energy.}
  \label{fig.11}
\end{figure*}

\begin{figure*}
 \centerline{\includegraphics[width=0.8\textwidth, clip=]{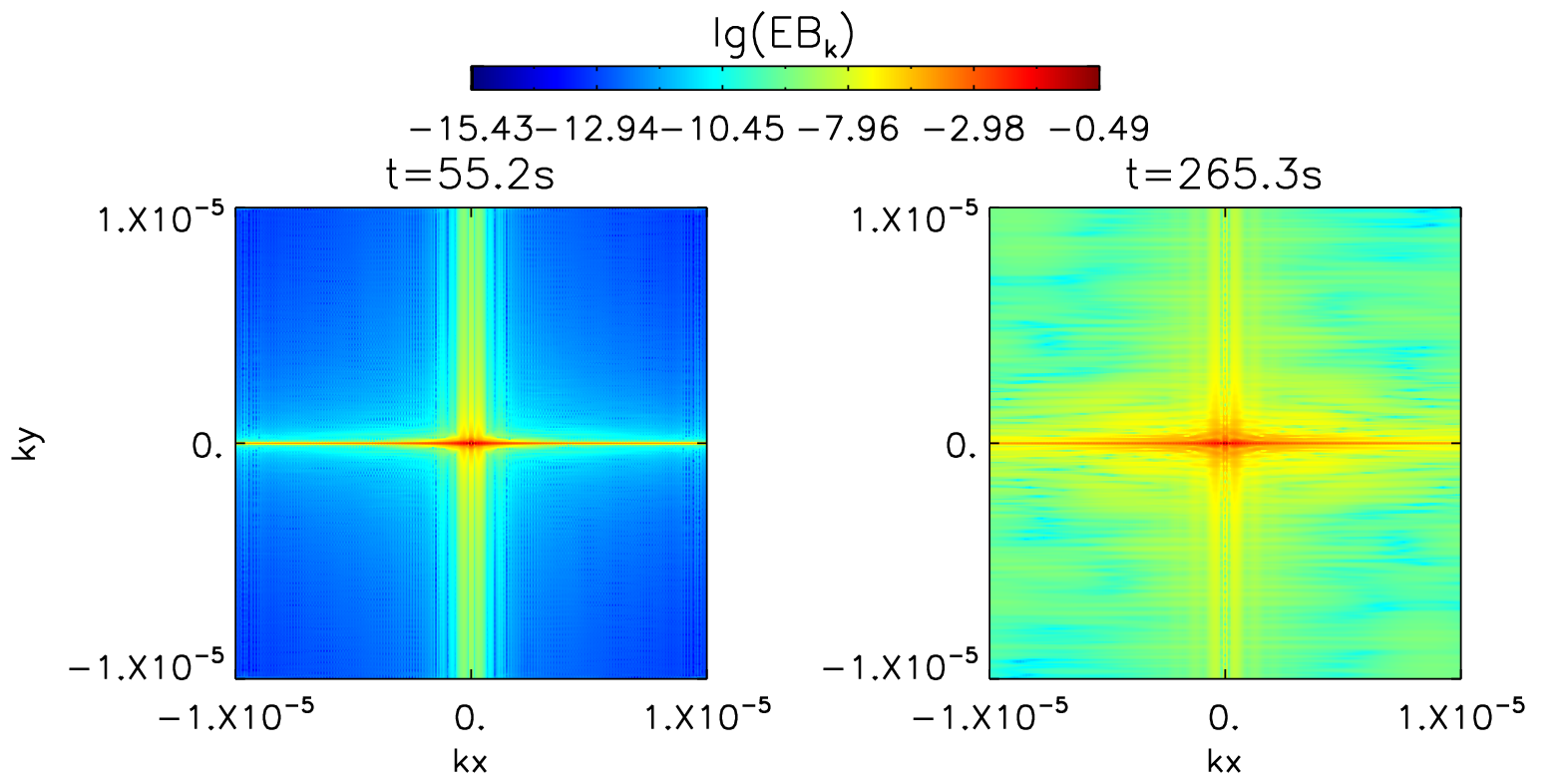}}
    \caption{The distribution of logarithmic values of magnetic energy in two-dimensional Fourier space at two different times.  }
  \label{fig.12}
\end{figure*}

\begin{figure*}
  \centerline{\includegraphics[width=0.8\textwidth, clip=]{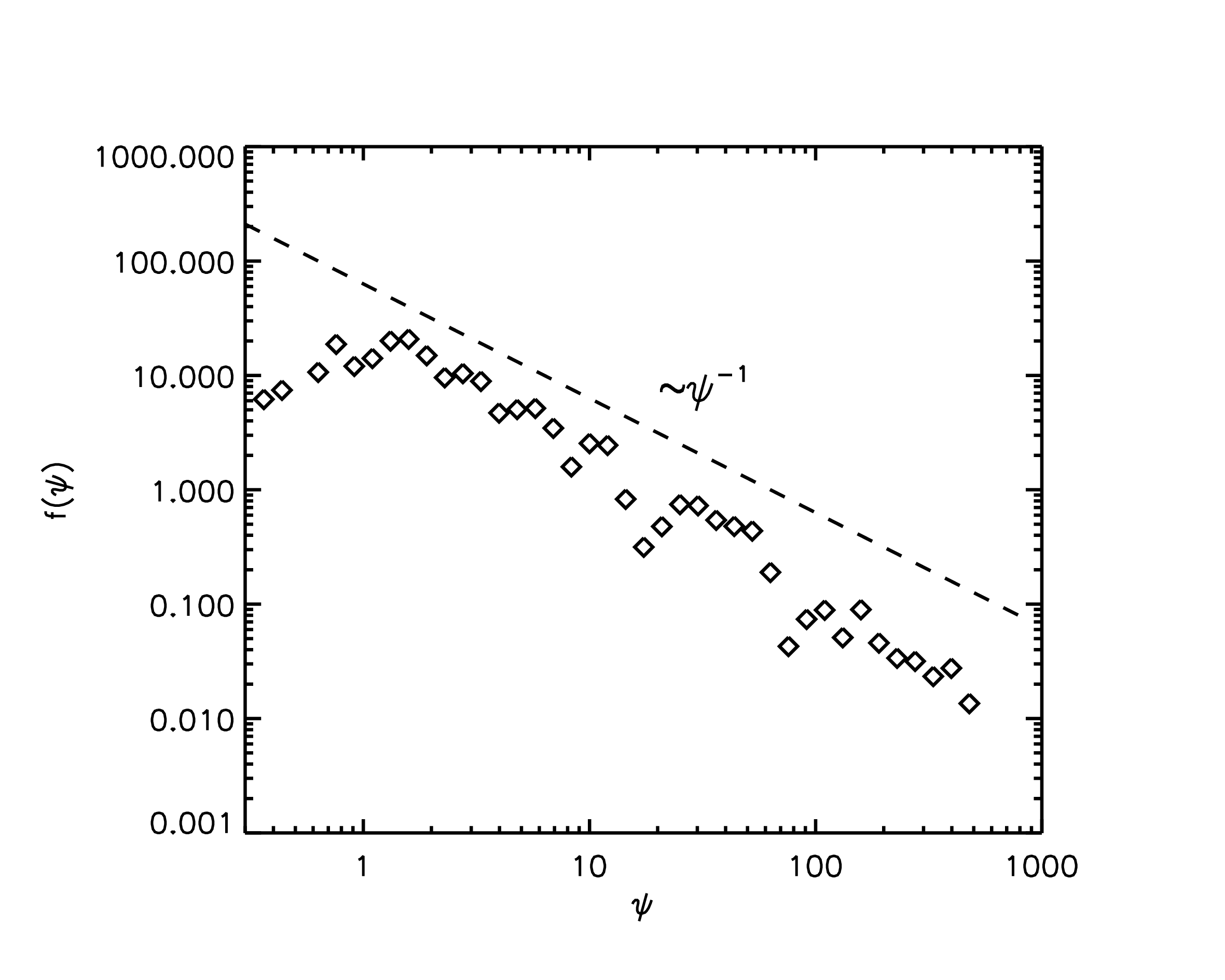}}
   \caption{The black diamond represents the plasmoid distribution function $f(\psi)$  and the dash straight line represents the pow law $\sim \psi^{-1}$. }
  \label{fig.13}
\end{figure*}

\end{document}